%% file: ms.tex
\documentclass[12pt,preprint]{aastex}

\shorttitle{Systematic Errors in SDSS Angular Clustering}
\shortauthors{Scranton et al.}

\begin{document}

\title{Analysis of Systematic Effects and Statistical Uncertainties in 
Angular Clustering of Galaxies from Early SDSS Data\altaffilmark{1}}

\author{Ryan Scranton\altaffilmark{2,3}, David Johnston\altaffilmark{2,3}, 
Scott Dodelson\altaffilmark{2,3}, Joshua A. Frieman\altaffilmark{2,3},
Andy Connolly\altaffilmark{4}, Daniel J. Eisenstein\altaffilmark{2,5,6}, 
James E. Gunn\altaffilmark{7}, Lam Hui\altaffilmark{8}, 
Bhuvnesh Jain\altaffilmark{9,10}, Stephen Kent\altaffilmark{3},
Jon Loveday\altaffilmark{11}, Vijay Narayanan\altaffilmark{7},
Robert C. Nichol\altaffilmark{12}, Liam O'Connell\altaffilmark{11},
Roman Scoccimarro\altaffilmark{13,14}, Ravi K. Sheth\altaffilmark{3}, 
Albert Stebbins\altaffilmark{3}, Michael A. Strauss\altaffilmark{7}, 
Alexander S. Szalay\altaffilmark{9}, Istv\'an Szapudi\altaffilmark{15}, 
Max Tegmark\altaffilmark{10}, Michael Vogeley\altaffilmark{16},
Idit Zehavi\altaffilmark{2,3}, James Annis\altaffilmark{3}, 
Neta A. Bahcall\altaffilmark{7}, Jon Brinkman\altaffilmark{17},
Istv\'an Csabai\altaffilmark{18,9},Robert Hindsley\altaffilmark{19},
Zeljko Ivezic\altaffilmark{7}, Rita S.J. Kim\altaffilmark{7}, 
Gillian R. Knapp\altaffilmark{7}, Don Q. Lamb\altaffilmark{2},
Brian C. Lee\altaffilmark{3}, Robert H. Lupton\altaffilmark{7}, 
Timothy McKay\altaffilmark{20}, Jeff Munn\altaffilmark{21}, 
John Peoples\altaffilmark{3}, Jeff Pier\altaffilmark{21}, Gordon T. 
Richards\altaffilmark{22}, Constance Rockosi\altaffilmark{2}, 
David Schlegel\altaffilmark{7}, Donald P. Schneider\altaffilmark{22},
Christopher Stoughton\altaffilmark{3}, Douglas L. Tucker\altaffilmark{3},
Brian Yanny\altaffilmark{3}, Donald G. York\altaffilmark{2,23}, for the 
SDSS Collaboration}

\altaffiltext{1}{Based on observations obtained with the Sloan Digital Sky
 Survey}
\altaffiltext{2}{Astronomy and Astrophysics Department, University of 
Chicago, Chicago, IL 60637, USA}
\altaffiltext{3}{NASA/Fermilab Astrophysics Center, P.O. Box 500, Batavia, IL 
60510, USA}
\altaffiltext{4}{University of Pittsburgh, Department of Physics and 
Astronomy, 3941 O'Hara Street, Pittsburgh, PA 15260, USA}
\altaffiltext{5} {Steward Observatory, University of Arizona, Tucson, AZ, 
85721, USA}
\altaffiltext{6}{Hubble Fellow}
\altaffiltext{7}{Princeton University Observatory, Princeton, NJ 08544, USA}
\altaffiltext{8}{Department of Physics, Columbia University, New York, NY 
10027, USA}
\altaffiltext{9}{Department of Physics and Astronomy, The Johns Hopkins 
University, 3701 San Martin Drive, Baltimore, MD 21218, USA}
\altaffiltext{10}{Department of Physics, University of Pennsylvania, 
Philadelphia, PA 19101, USA}
\altaffiltext{11}{Sussex Astronomy Centre, University of Sussex, Falmer, 
Brighton BN1 9QJ, UK}
\altaffiltext{12}{Department of Physics, 5000 Forbes Avenue, Carnegie Mellon 
University, Pittsburgh, PA 15213, USA}
\altaffiltext{13}{Institute for Advanced Study, School of Natural Sciences, 
Olden Lane, Princeton, NJ 08540, USA}
\altaffiltext{14}{Department of Physics, New York University, 4 Washington 
Place, New York, NY 10003}
\altaffiltext{15}{Institute for Astronomy, University of Hawaii, 2680 
Woodlawn Drive, Honolulu, HI 96822, USA}
\altaffiltext{16}{Department of Physics, Drexel University, Philadelphia, PA
19104, USA}
\altaffiltext{17}{Apache Point Obs., P.O. Box 59, Sunspot, NM 88349-0059}
\altaffiltext{18}{Department of Physics, E\"{o}tv\"{o}s University, Budapest,
Pf.\ 32, Hungary, H-1518}
\altaffiltext{19}{Remote Sensing Division, Code 7215, Naval Research 
Laboratory, 4555 Overlook Ave. SW, Washington, DC 20375}
\altaffiltext{20}{University of Michigan, Department of Physics, 
500 East University, Ann Arbor, MI 48109}
\altaffiltext{21}{U.S. Naval Observatory, Flagstaff Station, P.O. Box 1149,
Flagstaff, AZ, 86002-1149, USA}
\altaffiltext{22}{Department of Astronomy and Astrophysics, The Pennsylvania 
State University, University Park, PA, 16802, USA}
\altaffiltext{23}{Enrico Fermi Institute, 5640 S. Ellis Ave, Chicago, IL 60637,
USA}

\begin{abstract}

The angular distribution of galaxies encodes a wealth of information about 
large scale structure. Ultimately, the Sloan Digital Sky 
Survey (SDSS) will record the angular positions of order $10^8$ galaxies in 
five bands, adding significantly to the cosmological constraints. This is the 
first in a series of papers analyzing a rectangular stripe 
$2.5^\circ\times90^\circ$ from early SDSS data.  We present the angular 
correlation function for galaxies in four separate magnitude bins on angular 
scales ranging from $0.003$ degrees to $15$ degrees. Much of the focus of this 
paper is on potential systematic effects. We show that the final galaxy 
catalog -- with the mask accounting for regions of poor seeing, reddening, 
bright stars, etc. -- is free from external and internal systematic effects 
for galaxies brighter than $r^*= 22$. Our estimator of the angular 
correlation function includes the effects of the integral constraint and the 
mask. The full covariance matrix of errors in these estimates is derived using 
mock catalogs with further estimates using a number of other methods.

\end{abstract}


\section{Introduction}

One of the most direct and powerful probes of models of structure formation is
the two-point function for galaxies, either the correlation function in real 
space or the power spectrum in Fourier space. At least on large scales, 
observations of the power spectrum can be directly compared with predictions 
of theoretical models.  Indeed, this comparison is one of the strongest 
arguments (see e.g. Peacock \& Dodds, 1994) to date against the simplest Cold 
Dark Matter model with a matter density equal to the critical density. 

There are several ways to measure the power spectrum. The most direct is to 
use a redshift survey, which contains information not only about the two 
dimensional angular position of each galaxy but also about its radial 
distance from us.  Angular surveys do not have any radial information, but 
they are often just as powerful probes of the power spectrum because they 
contain many more galaxies than do redshift surveys. Examples of angular 
surveys which have been used to measure the power spectrum are the APM (Maddox 
et al., 1990; Efstathiou \& Moody, 2000) and the Edinburgh/Durham Southern 
Galaxy Catalogue (Collins, Nichol \& Lumsden, 1992; Huterer, Knox \& Nichol, 
2000). 

The Sloan Digital Sky Survey (SDSS; York et al. 2000; Gunn et al. 1998; 
Fukugita et al. 1996) will ultimately obtain angular positions for 
$\sim 10^8$ galaxies and redshifts for $10^6$ galaxies. Both will be powerful 
probes of cosmological models.  This paper analyzes the 
angular correlation function from early, imaging data taken during the  
photometric commissioning of SDSS. The survey data will be of higher quality 
(mainly due to better image quality and photometric calibration), so some of 
the systematic effects analyzed here will be less severe in the full survey.  
Likewise, since the data is collected digitally, we expect to be free of a 
number of systematic effects related to earlier surveys using scanned 
photographic plates (Nichol \& Collins, 1993; Maddox et al. 1996).

The data were taken in two nights in March, 1999 with the SDSS Camera (Gunn et 
al. 1998) on the 2.5m telescope (SDSS Runs 752/756).  The area surveyed is 
centered on the Celestial Equator $2.52$ degrees wide by approximately $90$ 
degrees long.  In equatorial coordinates, the observed region runs from 
$9^{h}40^{m}48^{s}$ to $15^{h}45^{m}12^{s}$ in $\alpha$ (J2000), putting the 
very ends of the run at somewhat low Galactic latitudes.  This area was 
imaged in two interlaced strips of six columns (or ``scanlines'') which 
together form a continuous region. Although each object is sampled in five 
bands ($u^\prime, g^\prime, r^\prime, i^\prime, z^\prime$; Fukugita et al. 
1996), the objects chosen here were selected based on their $r^*$ model 
magnitudes (where $r^*$ refers to the photometric calibration used in the 
Early Data Release (EDR; Stoughton, et al. (2001)) for the standard band-pass 
filter $r^\prime$).  Ultimately, photometric redshifts can be obtained by 
using the multi-band information, but here we make no estimate of the radial 
distance of each object.  This pair of runs has been used in previous early 
SDSS papers analyzing the galaxy luminosity function (Blanton et al. 2001), 
number counts (Yasuda et al. 2001) and colors (Shimasaku et al. 2001; 
Strateva et al. 2001).

Photometric calibration is carried out using an auxiliary 20''
telescope adjacent to the SDSS 2.5m telescope (the `Photometric
Telescope', or PT).  The PT observes a set of standard stars which have
been calibrated to the SDSS filter system (Smith et al. 2001) in order
to determine the atmospheric extinction of a given night.  Additionally, the PT
observes regions of the sky (`secondary standards') which overlap the
imaging scans, setting the photometric zeropoints for these.  For runs
752 and 756, 11 and 16 secondary patches were observed, respectively.  In 
later calibrations, the photometric zeropoint of each chip was assumed to be 
constant through each run, but the calibration of the data we used allowed 
for the zeropoint to vary from patch to patch.  The zeropoints from each 
secondary patch typically agreed with each other to 0.013 magnitudes in $r^*$.

Every object in the survey is assigned a probability that it is a galaxy
based upon its morphology.  The basic algorithm used to assign these 
probabilities is discussed in \S 2.  In \S 3, we test the star/galaxy 
separation scheme with a wide variety of systematic checks.  We show there 
that the separation predictably does not work well in regions of very poor 
seeing, so we mask out the poor seeing regions. The resultant mask is 
presented in \S 4; it accounts for seeing, reddening, bright stars, and 
saturated CCD columns. In \S 5, we look for systematic effects due to 
uncertainties in magnitudes.  Varying responses in different parts of the 
camera are another possible source of systematic errors, both within a given
scanline and from scanline to scanline.  We check for these in \S 6. 

The final third of the paper discusses technical details related specifically 
to the measurement of the angular correlation ($w(\theta)$).  Two estimators 
are used to estimate $w(\theta)$, one a point-based approach, the other 
cell-based. While they are equivalent on scales larger than a cell size, each 
carries with computational advantages and disadvantages.  These are discussed 
in \S 7, as is the integral constraint which becomes important on large scales 
(here on the order of a degree). The errors on $w(\theta)$ are particularly 
important because (i) they are due to both Poisson statistics and cosmic 
variance and (ii) they are highly correlated from bin to bin. We present 
estimates of the full covariance matrix in \S 8 using four techniques, each 
with its regime of validity.  Finally, we offer some conclusions in \S 9.

With this prescription for avoiding photometric systematic effects in 
hand, there are a number of clustering measurements possible.  A companion 
paper (Connolly et al. 2001) will present the final measurement of 
$w(\theta)$ along with comparisons to previously published measurements.  
Additionally, Tegmark et al. (2001) gives a measurement of the angular power 
spectrum ($C_l$).  Dodelson et al. (2001) inverts the angular correlations 
and angular power spectra to extract the three-dimensional power spectrum, 
which will then be used to do parameter estimation.  In parallel, Szalay et 
al. (2001) performs a Karhunen-Lo\'eve decomposition of the data, allowing for 
a direct estimation of the $\Gamma$ and $\sigma_8$ parameters.  Finally, 
Szapudi et al. (2001) presents the higher order correlation functions for the 
data.  All of these companion papers use a common data set, the EDR-P, taken
from the Early Data Release and extended to include the galaxy Probabilities
described below.

\section{Star/Galaxy Separation}

The photometric data are processed using a series of interlocking pipelines 
that flat-field the images, match up the data in the different bands, measure 
the properties of all detected objects, and apply astrometric and photometric 
calibrations.  A large number of attributes are measured for each object, 
including a variety of aperture and model magnitudes.  

Our object classification algorithm uses the outputs of this pipeline to 
separate stars and galaxies independent of the standard pipeline's binary 
decision about the stellar or galactic nature of a given object.  The 
pipeline's separation works well at relatively bright magnitudes where the 
distinction between galaxies and stars is clear-cut, but at the faint end of 
the magnitude range there is a definite need to know the degree of certainty 
in calling an object a star or a galaxy.  With that in mind, we developed a 
Bayesian method of star/galaxy separation based upon the outputs of the 
pipeline.  This method has proven effective enough that it will be a standard 
output of the future versions of the pipeline.  The details of this separation 
method are given in Lupton et al. (2001) along with more detailed descriptions 
of the processing pipeline and tests of the reliability of the morphological 
parameters.  For pedagogical purposes, we present an outline of the method 
employed for star/galaxy separation below.

	\subsection{Separation Method}\label{sec:sep_method}

	The data processing pipeline provides a number of standard outputs
which could be used for star/galaxy separation.  For our purposes, four 
magnitude measures are of principal interest: PSF magnitudes, exponential
magnitudes, deVaucouleurs magnitudes and model magnitudes.  The first of these
is simply the magnitude of a given object fit to the point-spread function 
(PSF) calculated locally based upon the measured PSF of nearby bright stars.  
The exponential and deVaucouleurs magnitudes are measured within two 
dimensional profiles where the axis ratio and scale lengths are fit to the 
object; in addition the model has been convolved with the PSF.  Model 
magnitudes are the best fit of either the exponential or deVaucouleurs model
in the $r^*$ band.

\begin{figure}

\plotone{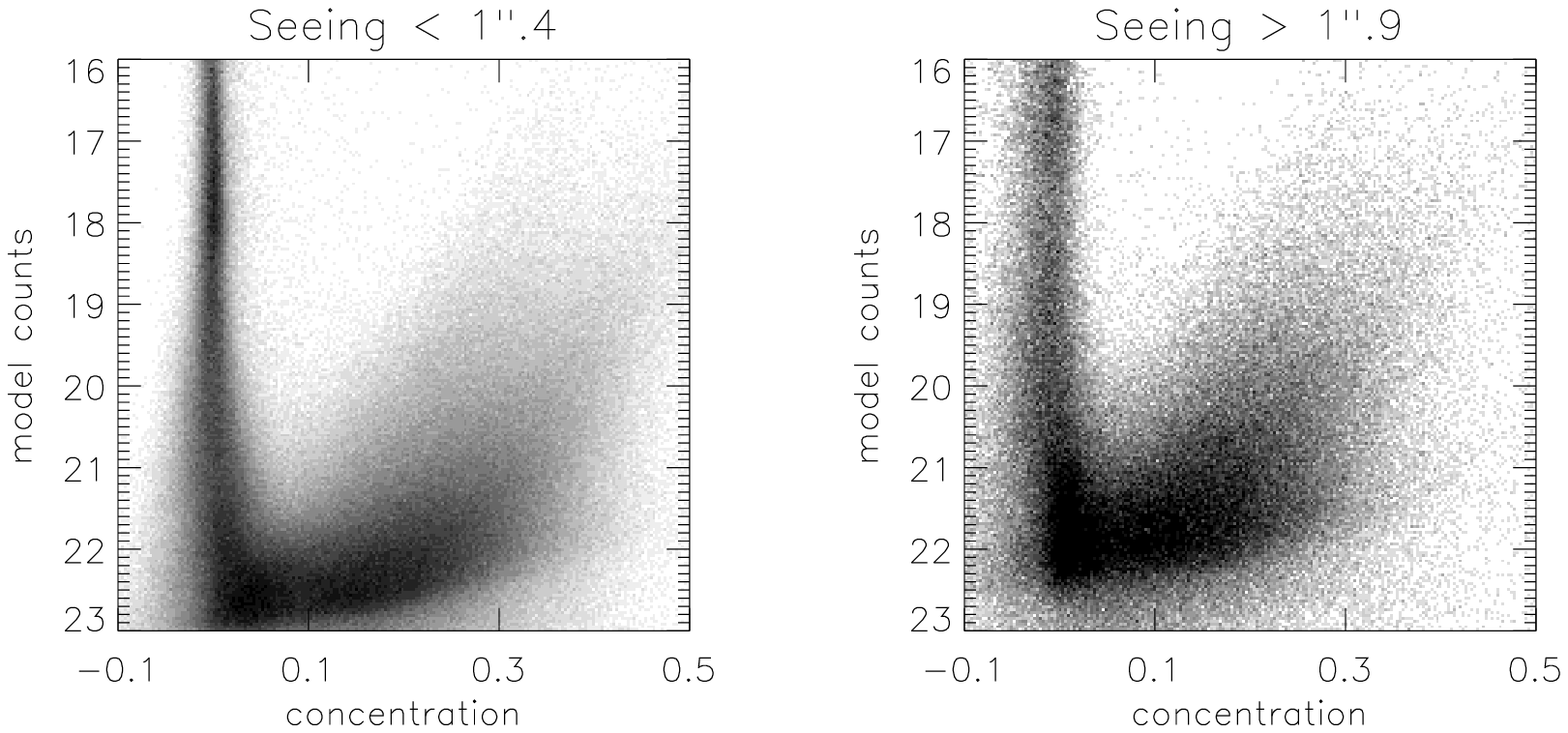}
\caption{Concentration-magnitude diagrams for regions of good (left) and bad
(right) seeing.  Areas of poor seeing show increased width of the stellar 
locus, brighter merging of the stellar and galactic loci and a shifting of the
faint galactic locus centroid toward that of the stellar locus.
\label{fig:see-con}}
\end{figure}

	From these magnitudes, we derive our central tool for star/galaxy 
separation, the {\it concentration}, which is defined for each object as 
$c \equiv r^*_{\rm PSF} - r^*_{\rm EXP}$, where $r^*_{\rm PSF}$ 
is the $r^*$ PSF magnitude and $r^*_{\rm EXP}$ is the exponential 
magnitude. In the case of a star, the concentration parameter should be very 
close to zero.  For a galaxy, however, the concentration is positive for 
bright magnitudes and then tends toward zero at fainter magnitudes as the 
galaxies become less and less resolved.  Figure~\ref{fig:see-con} shows the 
behavior of this parameter for several thousand objects over a range of model 
magnitudes in the $r^*$ band.

	The most striking feature of this plot is the clear 
separation between the stellar and galactic loci at bright magnitudes.  For
fainter magnitudes, this clean separation degrades as galaxies become 
less resolved and magnitude errors increase.  Clearly, this  will lead to some 
cross-contamination between the two populations, which we will quantify below.

\subsection{Bayesian Probabilities}

	While a straight-line binary cut in concentration-magnitude space 
(as given by the object type classification in the standard pipeline) has 
the advantage of simplicity (provided that one can adjust the location and
orientation of the demarcation line to maximize the selection efficiency), it 
produces little measure of the statistical confidence in the classification of 
each object.  The following gives a brief description of the probabilistic 
method of separation used in our analysis of $w(\theta)$.  Again, this method 
will be covered in greater detail in Lupton, et al. (2001).

Using the standard Bayesian formalism, we can express the probability that a 
given object is a galaxy (G) in terms of its magnitude ($m$) and concentration 
parameter ($c$) as 
\begin{equation}
P(G|m,c) = \frac{P(m,c|G) P(G)}{P(m,c)},
\end{equation}
where $P(m,c|G)$ is the posterior probability, $P(G)$ is the prior and $P(m,c)$
is the global likelihood. We can pull magnitude out of the expression for 
the posterior probability as 
\begin{equation}
P(G|m,c)  = \frac{P(c|m,G)P(m|G)P(G)}{P(m,c)},
\end{equation}
where $P(m|G)$ is simply the galaxy number count relation for a given 
magnitude.  We can find this by using a simple straight line cut for the 
brighter magnitude objects (approximately $17 \le r^* \le 19$) where the 
stellar and galactic populations are well separated.  We can fit an 
exponential curve to this relation and normalize it over the magnitude range 
to find the probability for a given magnitude.  Yasuda et al. (2001) have 
measured this relation from the same SDSS data; our independent measurement 
confirms their result for the fainter end of their sample.  

\begin{figure}
\plotone{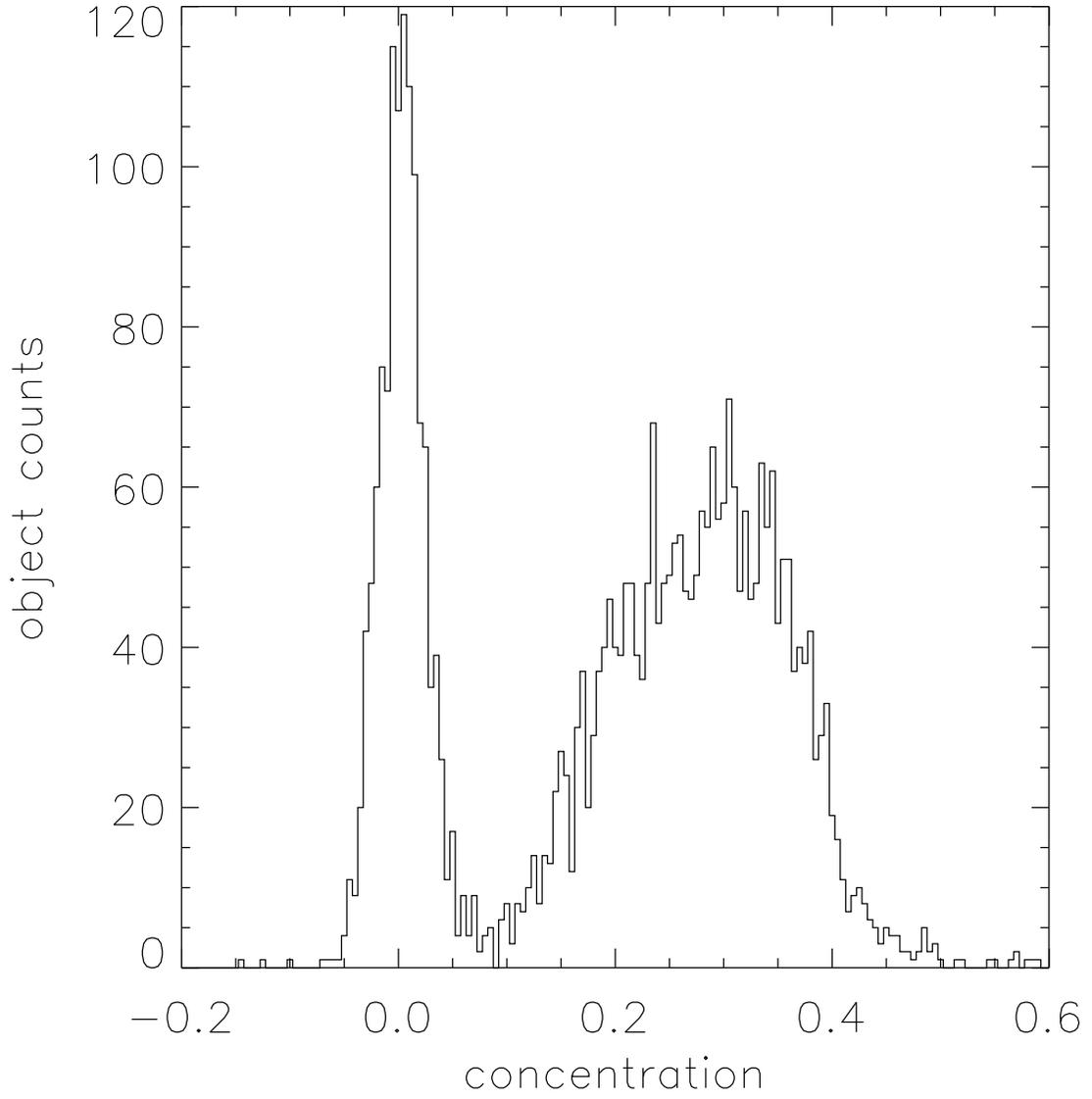}
\caption{Concentration histogram for a section of run 756 for objects with
$20.0 \le r^* \le 20.5$.\label{fig:con_hist}}
\end{figure}

	Using the fact that the stellar and galactic probabilities for a given 
object must sum to one, we can re-write the above as 
\begin{equation}
P(G|m,c) = \left ( 1 + \frac{P(c|m,S) P(m|S)P(S)}{P(c|m,G)
P(m|G)P(G)} \right )^{-1},
\end{equation}
Again, we can use the empirical results from the easily separable bright 
objects to find $P(m|G)$ and $P(m|S)$, taking into account the variation in 
stellar density as a function of Galactic coordinates.  In practice, the 
sensitivity of the separation to the star-galaxy ratio is generally quite low 
for realistic values.  More explicitly, since we will be calculating 
these quantities over a small, finite magnitude range, we can re-express the
above as
\begin{equation}
P(G|m,c) = \left ( 1 + \frac{P(c|m,S)P(S|m)}{P(c|m,G)P(G|m)} \right )^{-1},
\label{eq:no-see_prob}
\end{equation}
where $P(S|m)$ and $P(G|m)$ folds in the relative abundance of galaxies and 
stars due to changes in Galactic latitude. This leaves us only 
$P(c|m,S)$ and $P(c|m,G)$ to calculate.  To find these probabilities, we bin 
the data from the magnitude-concentration plot in magnitude, resulting in 
histograms like that in Figure~\ref{fig:con_hist}.  After applying a simple 
transformation on the concentration to rein in the tail on the galaxy 
distribution, we fit a Gaussian to the galaxy locus and two Gaussians to the 
stellar locus (to account for the slightly wider non-Gaussian tails) for all 
of the magnitude bins.  This allows us to interpolate the parameters of the 
two probability distributions, giving us the galactic and stellar 
probabilities for a given magnitude and concentration.  

	With minimal effort, we can expand the above to include information on 
the seeing conditions ($s$) for a given object, resulting in 
\begin{equation}
P(G|m,c,s) = \left ( 1 + \frac{P(c|s,m,S) P(S|m,s)}
{P(c|s,m,G) P(G|m,s)} \right )^{-1}.\label{eq:not_final_prob}
\end{equation}
This extension is needed to compensate for the different behavior of the 
stellar and galactic loci under different seeing conditions as seen in 
Figure~\ref{fig:see-con}.  In regions where the seeing is very good, there
is clear separation between the stellar and galactic loci to fainter magnitudes
than in those regions with poor seeing.  Likewise, the centroid of the 
galactic locus is considerably closer to that of the much wider stellar locus 
at fainter magnitudes in the bad seeing regions, where the PSF has increased.

In modifying Equation~\ref{eq:no-see_prob} to include seeing in this way, we 
are assuming that the measurement of the magnitude is unaffected by seeing.  
For brighter objects this should be true and in the faint limit the effects of 
the seeing on the magnitude would act in much the same manner for both 
galaxies and stars since both types of object have nearly the same light 
distribution.  This effect should therefore roughly cancel in 
Equation~\ref{eq:not_final_prob}.  This conjecture has been verified by 
Ivezic et al. (2001) in their examination of objects doubly imaged 
in those regions where scanlines in interlaced strips overlap (see York et al.
(2000) for an explanation of the scanning procedures).  They have 
found that, for reasonably bright objects imaged in very different seeing 
conditions, the magnitudes are very consistent.  Just as importantly, 
the effect of different seeing on the magnitude was the same for stars and 
galaxies at the faint limit.  Thus, we can safely bin the objects in 
both magnitude and seeing before fitting the Gaussians to the concentration 
distributions and then bilinearly interpolate in those variables to find 
$P(c|s,m,G)$ for a given concentration, magnitude, and seeing.  

	For the actual form of $P(G|s,m)$ and $P(S|s,m)$, we can use a 
similar method to that used for the case without seeing included.  The effect 
of worsening seeing is to brighten the faint magnitude limit.  Since most of 
the objects at that limit have similar sizes anyway, we would again expect 
that the effects for galaxies and stars in that limit would be the same and 
thus cancel out in the formula.  In fact, we can replace $P(G|m,s)$ with 
$P(G|m)$ and $P(S|m,s)$ with $P(S|m)$ without losing information,
\begin{equation}
P(G|m,c,s) = \left ( 1 + \frac{P(c|s,m,S) P(S|m)}
{P(c|s,m,G) P(G|m)} \right )^{-1}.\label{eq:final_prob}
\end{equation}

	With a star/galaxy separation scheme in hand, we must verify that 
applying the method results in a uniform sample of galaxies across the field 
of view.  This will make checking the variation of the sample against 
possible sources of contamination paramount if we want to assure ourselves we 
are measuring the galaxy clustering independent of systematic 
effects.  As we will show in the following sections, the proper cuts on those
systematic contaminants allow us to reliably separate stars from galaxies
down to a model magnitude of 22 in $r^{*}$.  The efficacy of the original 
binary galaxy/separation separation has been analyzed by Yasuda et al. (2001) 
down to $r^* \sim 21$.  Our tests verify that our probabilistic separation 
matches this performance and allows us to go to fainter magnitudes where the 
binary method fails.  This allows us to make the four unit magnitude cuts that 
we will use for the rest of our analysis: $18 \le r^* \le 19$, 
$19 \le r^* \le 20$, $20 \le r^* \le 21$, and 
$21 \le r^* \le 22$, with approximately 0.16, 0.31, 0.65 and 1.15 million 
galaxies, respectively.  All of the magnitude cuts are based on the model
magnitudes, dereddened using the reddening map of Schlegel, Finkbeiner, \& 
Davis (SFD, 1998).

\section{External Systematic Error Sources} \label{sec:separation}

	By restricting the area of our survey, we can reduce the measured 
systematic errors in $w(\theta)$ due to variations in seeing and dust 
extinction to below the errors in the measurement due to cosmic variance and 
Poisson errors.  We are also able to separate the stellar and galaxy 
populations to the extent that their cross-contamination becomes negligible.
In this section, we present various diagnostic tests of the data to determine
how best to define the survey area.  Although these tests concentrate on the 
angular correlations, the resulting mask is equally valid for any measurement
of angular clustering (e.g. angular power spectrum or KL decomposition).

	\subsection{Galaxy Densities}

\begin{figure}
\plottwo{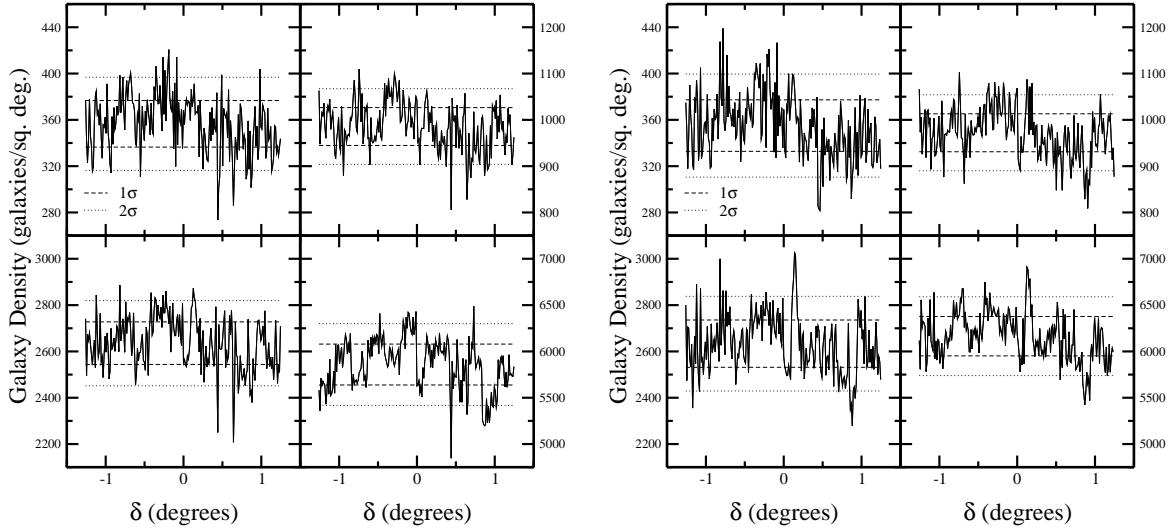}{f4.eps}
\caption{Galaxy density projected along the long axis of the survey area 
without using a mask (left) and using a mask (right).  Clockwise from the upper
left, the magnitude bins are $18 \le r^* \le 19$, 
$19 \le r^* \le 20$, $20 \le r^* \le 21$, $21 \le r^* \le 22$
.  Horizontal lines indicate $1\sigma$ and $2\sigma$ limits in the density.
\label{fig:dec_density}}
\end{figure}

	Measuring the galaxy densities projected along the long and short 
axes of the survey area is the first check that our galaxy sample is 
reasonably uniform.  The first check is verifying that the structure of the 
scanlines is not reflected in the galaxy densities.  The left panel of 
Figure~\ref{fig:dec_density} shows the variation of the galaxy density in the 
raw data for each of the magnitude bins as a function of declination, with 
CCD scanlines progressing from left to right; alternating scanlines are 
observed simultaneously.  The width of each scanline is $\sim 0.21^\circ$ and 
the 12 scanlines are split evenly between positive and negative declination.

Here we obtain the galaxy densities by summing the galaxy probabilities of 
all objects in the region. To first order, the appearance of boundaries 
between scanlines is minimal, but certainly visible in the break at zero 
declination, for example.  If we apply a mask to the data (the exact details 
of which are explained and justified in \S\ref{sec:cross}) to avoid the 
regions where the data quality is questionable, we get the result plotted 
in the right panel of Figure~\ref{fig:dec_density}.  The masked sample avoids 
obvious problems like the sharp dips near $\delta \sim 0.4^\circ$ and 
$0.65^\circ$ (explained in \S\ref{sec:mask}) and the sharp boundaries between
scanlines.

	In order to get a more precise idea of what data should be cut out if 
we want to avoid systematic errors, we need to examine the behavior of the 
galaxy density while varying some of the possible sources of errors.  Once 
this is done, we will be able to move onto more sophisticate techniques for
determining the observational limits on these external sources.

		\subsection{Seeing Variations}
	
	As part of the photometric pipeline, a set of PSF eigencomponents are 
determined using Karhunen-Lo\'eve decomposition of the bright stars in each 
field of each scanline (the details of this process are presented in 
Lupton et al. (2001)).  By taking into account the position of these stars, 
one can use interpolation to reconstruct the PSF from a combination of these 
eigencomponents for any object in the field.  To determine the seeing for each 
object, we calculated the second moment of each of these eigencomponents and 
then used the same interpolation scheme to reconstruct the seeing; the seeing
is given as 2.355 times the second moment which would give the FWHM under
the assumption that the PSF is Gaussian.  This allows the seeing to be 
calculated at any object without the more time consuming process of 
re-constructing the PSF at that point and calculating its second moment.  It 
should be noted, however, that this definition differs from that used by 
Yasuda et al. (2001) in their analysis, resulting in qualitative, but not 
quantitative, agreement between our seeing and theirs.

\begin{figure}
\plottwo{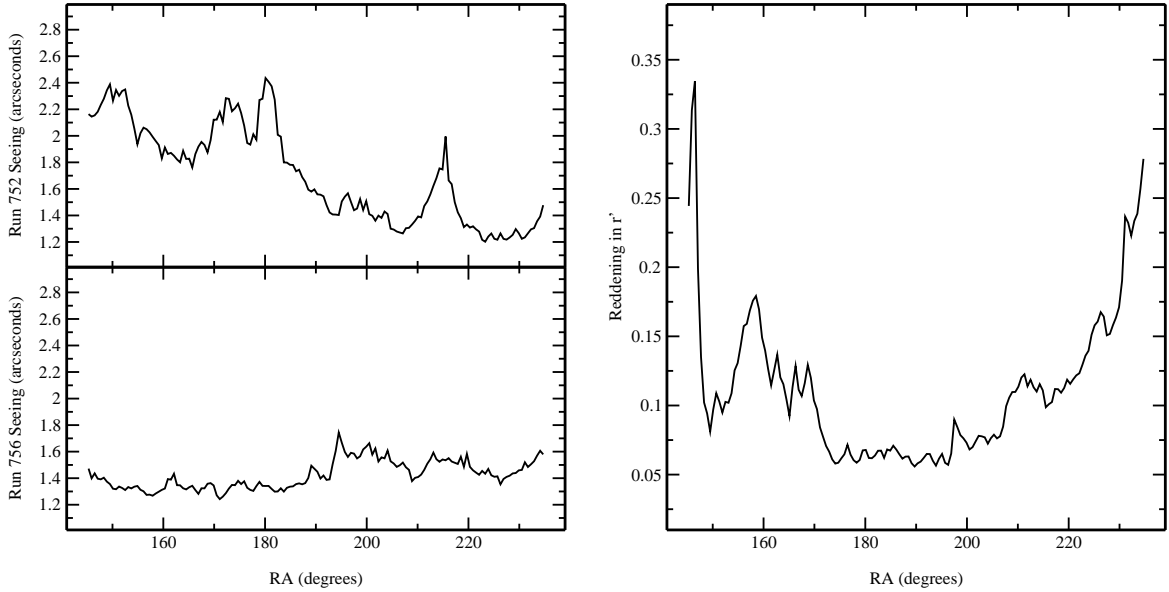}{f6.eps}
\caption{Seeing variations for each of the strips as a function of $\alpha$ 
(left) and reddening variations as a function of $\alpha$ (right).
\label{fig:ra_var}}
\end{figure}

	The left panel of Figure~\ref{fig:ra_var} shows the mean seeing 
variation across the scanlines for the two runs as a function of right 
ascension.  The seeing for Run 756 is generally well-behaved throughout the 
course of the run, with only the occasional departure above $1''.6$.  
Run 752, taken two days prior to Run 756, is much more volatile; the entire 
first half of the run oscillates above $1''.8$ and then later the seeing 
spikes to $2''.0$.  This structure in the seeing map will require extensive 
masking and require careful checking against false signal on the scanline 
scale.

\begin{figure}
\plottwo{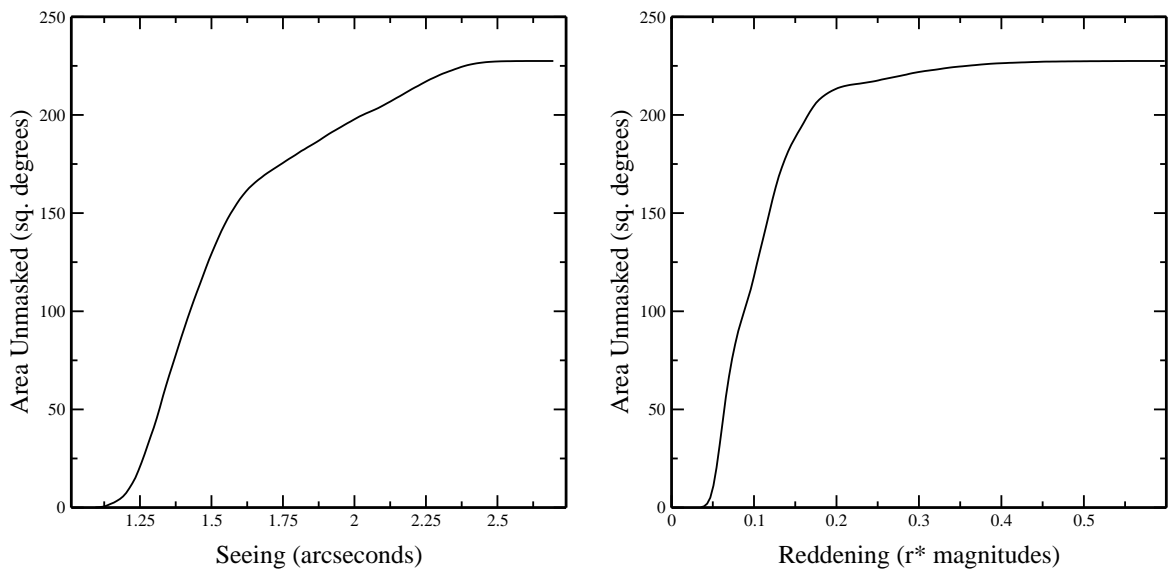}{f10.eps}
\caption{Area left unmasked versus seeing requirement (left) and reddening
requirement (right). \label{fig:bin_area}}
\end{figure}

In normal survey operations, regions where the seeing degraded to 
worse than $1''.5$ are marked for re-observation, but we did not have 
that luxury for the commissioning data.  In the left panel of 
Figure~\ref{fig:see-red_bin}, we plot the mean galaxy density for the combined 
stripe as a function of seeing.  The galaxy density varies considerably for 
the faintest magnitude bin over the factor of two in seeing conditions, 
suggesting that poor seeing lowers the confidence that a given object is a 
galaxy.  Not surprisingly, we also see that the effect of poor seeing is more 
pronounced at fainter magnitudes than for the brighter objects.  Already, the 
magnitude bin from $21 \le r^* \le 22$ clearly shows that we need to 
restrict the data to seeing better than $1''.75$, but the cross-correlation 
analysis below will show that the cut needs to be even more restrictive.  
Figure~\ref{fig:bin_area} shows the area which would remain unmasked for a 
given seeing cut.

	\subsection{Reddening Variations}

	Since we are only analyzing the effect of intermediary dust in 
a single band, the term ``reddening'' is not as appropriate as ``extinction''
or ``absorption''.  However, the magnitude extinction limits that we will set 
in constructing our mask will refer to the $r^*$ element of the 
{\tt reddening} output of the photometric pipeline, so we will adopt the use
of ``reddening'' in favor of other alternatives to avoid confusion.

	The right panels of Figures~\ref{fig:ra_var} and \ref{fig:see-red_bin} 
repeat the above analysis in terms of the SFD reddening.  The dependence of 
galaxy density on reddening is weaker than for seeing, but that is to be 
expected since the survey area does not contain much area where the reddening 
is significantly higher than $0.2$ magnitudes.  Likewise, the small fraction 
of the area with reddening less than $0.05$ magnitudes makes that density 
measurement highly dependent on large-scale structure variations in those 
regions.  Still, the fact that the scatter in density is so much larger than 
the Poisson error for those higher reddening areas suggests that we should 
consider setting the limit for reddening around $0.2$ magnitudes.  
Figure~\ref{fig:bin_area} shows that the area excluded by such a cut is small.

	\subsection{Cross-Correlations} \label{sec:cross}

	Cross-correlations of the galaxy density with maps of external sources
offer the most powerful means for checking against contamination in the galaxy 
sample.  Not only can they detect systematic effects, they also offer 
information on the angular scale of that correlation.  This is particularly 
important in the case of seeing, where we have sharp variations between 
adjacent scanlines.  The caveat with such an approach is the unstated 
assumption that any variation in the cross-correlations is due to an external
source, rather than a variation in the observing system.  As we will see in 
\S\ref{sec:internal}, this is a reasonable enough assumption and so we will 
proceed by cross-correlating the galaxy density with seeing, reddening, 
stellar density and sky brightness.

	To measure the cross-correlations, we generated a pixelized version of
the data, breaking the area in each $0.21^\circ$ wide scanline into square 
cells approximately $0.04^\circ$ on a side.  This gives us five cells in the 
$\delta$ direction for each scanline and approximately 10,000 for the whole 
of a given scanline.  In each cell, we find the mean seeing, mean 
reddening in $r^*$ and mean sky brightness in $r^*$ for all of the 
objects in the cell, as well as the sum of the galaxy and star 
likelihoods in each of the four magnitude bins.  In principle, these 
quantities could be found using the seeing, reddening and sky brightness maps 
independently, but this measure weights our average toward the values most 
relevant to the objects in the cell.

	The size of the cells ensures that the majority of the cells will 
contain on order 30 objects down to $r^*=22$.  Smaller cells would 
allow for greater resolution, but we suspect that most of the systematic 
effects will occur on the scale size of the scanline or larger.  The cell size 
is also of order the angular resolution of the SFD reddening map.  Keeping the 
mean number of objects per cell high also allows us to ignore cells without 
any objects (usually due to a missing area in the data, as happens with a 
single irreducible field in scanline 4 of run 752) without biasing ourselves 
against genuine voids.  

	Having the information in this form allows us to calculate 
the cross-correlation of the galaxy catalog with seeing, reddening, stellar 
density and sky brightness in each of the magnitude bins for a variety of 
different seeing and reddening cuts.  Once we have established the limits on 
seeing and reddening necessary to ensure an uncontaminated sample, we then 
use this pixelization to construct the mask. 

\begin{figure}
\plottwo{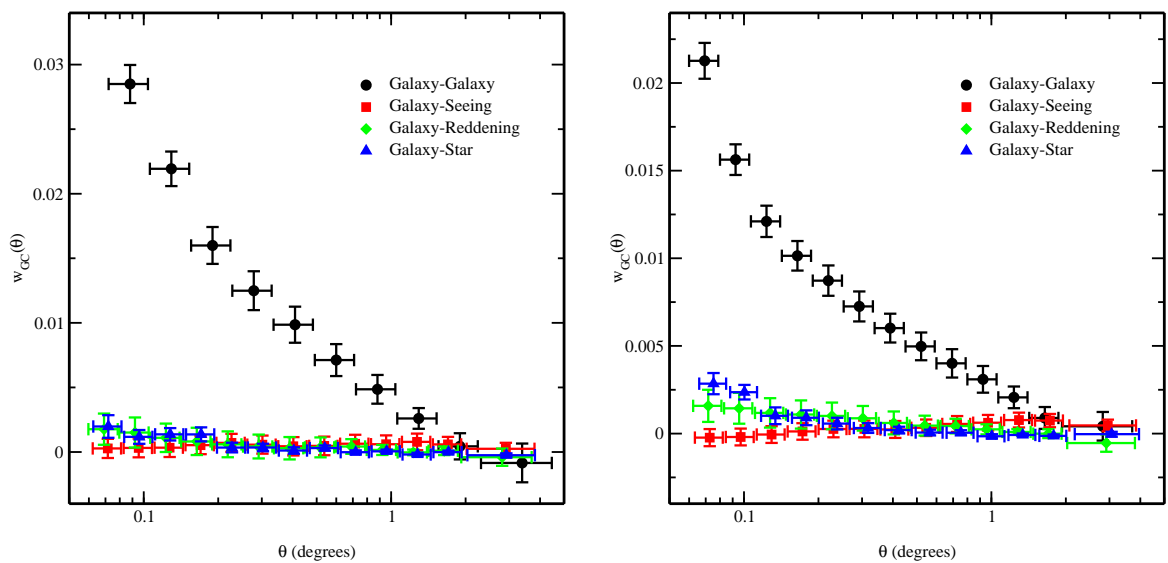}{f18.eps}
\caption{Comparison of the galaxy auto-correlation to the cross-correlation of
the galaxy density with seeing, reddening and stellar density, respectively, 
for the magnitude bins $20 \le r^* \le 21$ (left) and 
$21 \le r^* \le 22$ (right). 
\label{fig:all_corr_small}}
\end{figure}

	To measure the cross-correlation, we first divide the stripe into 35
separate square regions, approximately 2.5 degrees on a side, each containing
$\sim 3600$ cells.  For each square, we calculate the mean sum of galaxy 
probabilities ($\bar{n}^g$) per cell in a given magnitude bin, as well as the 
mean for the possible contaminant ($\bar{x}^c$), where $x^c$ could refer to 
the sum of the stellar probabilities, mean seeing, etc.  This allows us to 
calculate the fractional galaxy and contaminant overdensity in a given cell 
$i$,
\begin{eqnarray}
\delta^g_i &=& \frac{n^g_i - \bar{n}^g}{\bar{n}^g} \label{eq:delta_def} \\
\nonumber \delta^c_i &=& \frac{x^c_i - \bar{x}^c}{\bar{x}^c} \label{eq:contam}
\end{eqnarray}
The cross-correlation, $w_{gc}(\theta)$, is then simply
\begin{equation}
w_{gc}(\theta_\alpha) = 
\frac{\sum_{i,j} \delta^g_i \delta^c_j \Theta^\alpha_{ij}}
{\sum_{i^*,j^*}\Theta^\alpha_{i^*,j^*}},
\end{equation}
where $\Theta^\alpha_{ij}$ is unity if the separation between cells $i$ and $j$
is within angular bin $\theta_\alpha$ and zero otherwise.  Once the 
measurement has been done in each of the 35 sub-samples, we calculate the mean 
($\bar{w}_{gc}(\theta)$) and error on the mean ($\Delta\bar{w}_{gc}(\theta)$),
\begin{equation}
(\Delta\bar{w}_{gc}(\theta))^2 = \frac{1}{N^2}\sum^N_{i=1}
(\bar{w}_{gc}(\theta) - w_{gc,i}(\theta))^{2}, \label{eq:sub-error}
\end{equation}
where $N=35$ in this case.  In examining the cross-correlations below, one 
should bear in mind that the galaxy auto-correlation signal we observe at 1 
degree is approximately $0.005 \pm 0.0025$ for the faintest magnitude bin
(see Figure~\ref{fig:all_corr_small}).  Cross-correlation signals smaller than
this value will be over-whelmed by the galaxy signal.

		\subsubsection{Seeing} \label{sec:see}

\begin{figure}
\plottwo{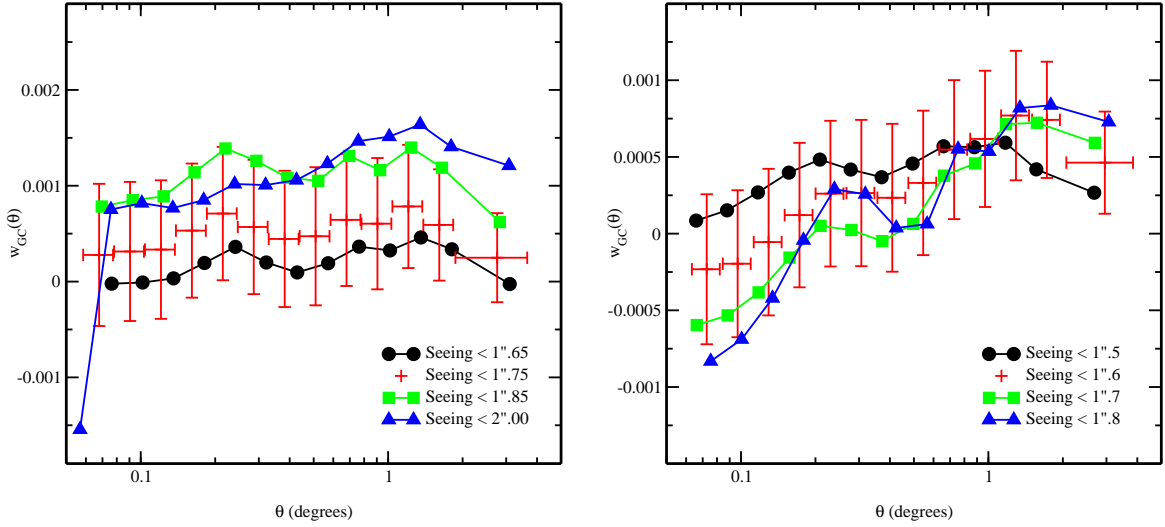}{f12.eps}
\caption{Galaxy-Seeing cross-correlations for magnitude bins $20 \le r^*
\le 21$ (left) and $21 \le r^* \le 22$ (right).  The size of the error 
bars on the preferred seeing cut are typical of those for the other seeing 
cuts, which have been eliminated for clarity.\label{fig:gal-see}}
\end{figure}

	The cross-correlations between the seeing and the galaxy density for 
the faintest two magnitude bins in the sample ($20 \le r^* \le 21$ and 
$21 \le r^* \le 22$) are shown in Figure~\ref{fig:gal-see}.  The goal 
here should be a flat curve, consistent with zero, and in particular one 
that shows no structure on the $0.21$ degree scale of the scanlines.  
Such structure is still seen with a 1''.7 seeing cut for the faintest 
magnitude bin.  The cross-correlation signal is reduced to an acceptable 
level by using a cut at $1''.6$ seeing. It should be noted that even 
the slight departure from zero seen with this cut is still well below the 
measurement of $w(\theta)$ on the same angular scales 
(Figure~\ref{fig:all_corr_small}).

	Making a cut at $1''.6$ is necessary for the faintest bin, 
but if we are interested in brighter objects, we can relax this cut somewhat.  
The left panel of Figure~\ref{fig:gal-see} shows the same galaxy-seeing 
cross-correlation for objects with magnitudes $20 \le r^* \le  21$.  In 
this case, we see that we can raise the seeing limit to $1''.75$ and 
still have a cross-correlation consistent with zero, although with some slight 
variation on the scale of the scanlines.  Since we want to include as much 
area as possible, we use two cuts, one for the faintest bin cutting at seeing 
of $1''.6$ and a second excluding seeing worse than $1.''75$ to use for 
the other three brighter magnitude bins. 

		\subsubsection{Reddening} \label{sec:red}

\begin{figure}
\plottwo{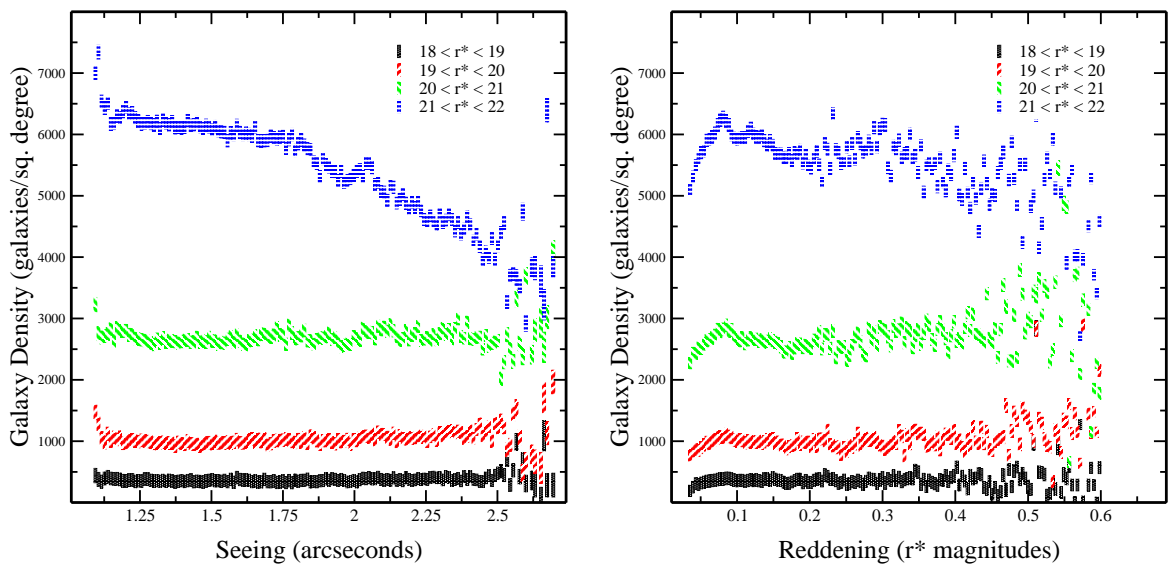}{f8.eps}
\caption{Galaxy density for four magnitude bins as a function of the local 
seeing (left) and reddening (right).\label{fig:see-red_bin}}
\end{figure}

\begin{figure}
\plottwo{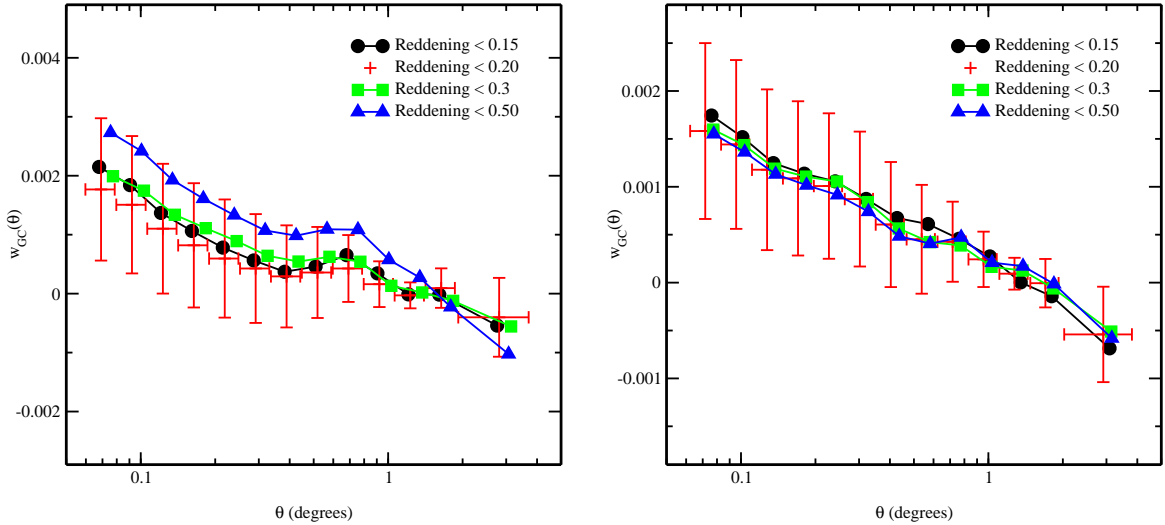}{f14.eps}
\caption{Galaxy-Reddening cross-correlations for magnitude bins 
$20 \le r^* \le 21$ (left) and $21 \le r^* \le 22$ (right).  Again,
the error bars on the favored cut are typical of the other limits.
\label{fig:gal-dust}}
\end{figure}

	Unlike the galaxy-seeing cross-correlations, there is not a strong 
relation between tightening the restriction on the allowed reddening in the
$r^*$ band and improved lack of cross-correlation 
(Figure~\ref{fig:gal-dust}).  This is not terribly surprising given the 
fluctuations in galaxy density as a function of reddening we saw in 
Figure~\ref{fig:see-red_bin}, particularly for the fainter magnitude bins.  
However, while the cross-correlations show some degree of angular dependence, 
they are easily within $2\sigma$ of zero for all angular scales and below the 
level of the galaxy auto-correlation errors.

	Given this, we exclude those regions where the reddening is worse than
$0.2$ magnitudes in $r^*$ as is suggested by the scatter in galaxy 
densities at higher reddening levels in Figure~\ref{fig:see-red_bin}.

		\subsubsection{Stellar Density}

\begin{figure}
\plottwo{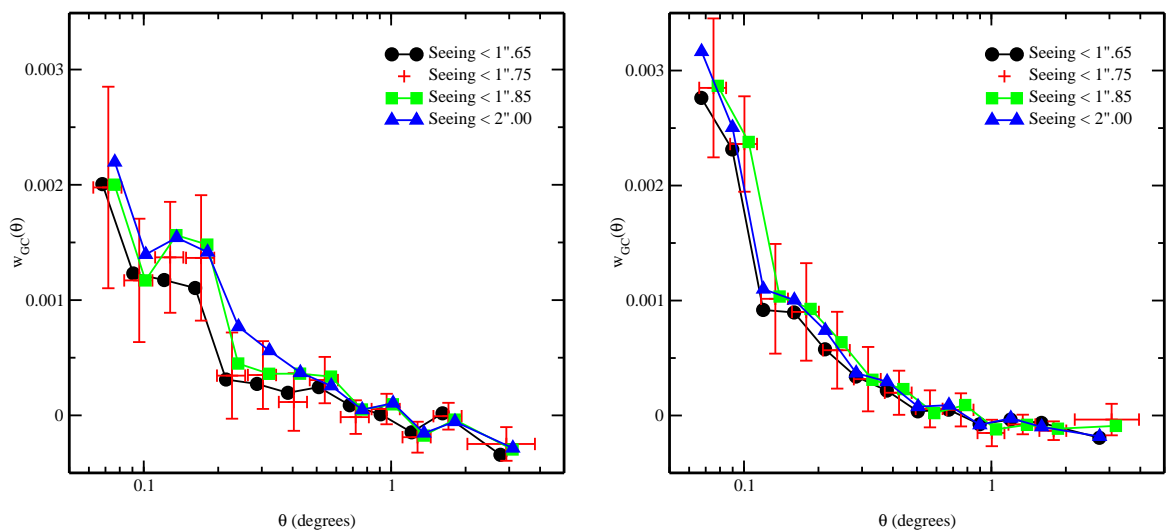}{f16.eps}
\caption{Galaxy-Star cross-correlations for magnitude bins $20 \le r^*
\le 21$ (left) and $21 \le r^* \le 22$ (right).  The error bars on the
favored limit are typical of the other limits.\label{fig:gal-star}}
\end{figure}

With perfect star/galaxy separation, we would expect the stellar density 
auto-correlation to be consistent with zero, except perhaps on the very 
smallest scales.  Thus, in the case where we have mistaken galaxies for stars 
and vice-versa, we would expect that the cross-correlation of these samples 
would produce a damped version of the galaxy auto-correlation, diluted by the
effectively null stellar auto-correlation.  For these tests, we use the limits 
on the seeing (better than $1''.6$ for galaxies fainter than 
$r^* = 21$ and better than $1''.75$ for brighter galaxies) and 
reddening (reddening less than 0.2 in $r^*$) established in 
\S\ref{sec:see} and \S\ref{sec:red}.  As shown in Figure~\ref{fig:gal-star}, 
the cross-correlation between the galactic and stellar populations is within 
the $2\sigma$ limit of zero for magnitudes brighter than $r^* \sim 21$.  For 
the faintest magnitude bin, however, there is a definite correlation between 
the two populations at small angles, again most likely due to some leakage 
between the two samples.  However, we can see from 
Figure~\ref{fig:all_corr_small} that the deviation from zero for 
the star-galaxy cross-correlation is not only much less than the galaxy-galaxy 
auto-correlation itself , but is also less than the error on that measurement 
even for the faintest magnitude bin.

		\subsubsection{Sky Brightness}

	We also consider the cross-correlation between the galaxy density
and the sky brightness.  Since our faintest two bins approach the limit of 
the photometric system (York et al. 2000), we might expect that the confusion
between a fluctuating sky brightness and the outer edges of galaxies might
result in an anti-correlation of sky brightness and galaxy density.  As 
expected, there is a slight ($\sim 10^{-5}$ with $50\%$ error), but non-zero, 
anti-correlation on the smallest angular scales.  However, the amplitude of 
this cross-correlation is well below the level of the errors on $w(\theta)$ 
(Figure~\ref{fig:all_corr_small}).

		\subsubsection{Large Angle Cross-Correlations} 
\label{sec:large-scale}

	Finally, we need to consider the large angle effects of variations in 
reddening and seeing.  Our previous calculations were primarily concerned with
the effect of these variations on the scale size of the scanlines, where we
expected to see discontinuities in the seeing.  While eliminating 
cross-correlations on that scale is important, it does not guarantee that we
do not have larger scale cross-correlations which could cause problems for the
$C_l$ and KL measurements of the data in Tegmark et al. (2001) and 
Szalay et al. (2001).  
	
	Unlike the smaller angle measurements, the sub-sampling method
is not appropriate for calculating the error on this measurement.  Rather, we
use a variation on the jackknife error scheme, allowing us to use the whole
data set.  For a traditional jackknife, we would perform the measurement $N$ 
times, removing a single different data point each time.  In our form, we use 
sub-samples similar (but not identical) to those described in 
\S\ref{sec:cross} as our unit of subtraction, calculating the galaxy 
auto-correlation $N$ times, each time excluding a different sub-sample.  To 
ensure that we have enough measurements to constrain the 23 angular bins for 
this measurement, we used 26 samplings of the data.  In this scheme, the error 
is given as  
\begin{equation}    
(\Delta w(\theta))^2 = \frac{N-1}{N} \sum^N_{i=1} 
(\bar{w}(\theta) - w_{i}(\theta))^2, 
\label{eq:jack-error}
\end{equation}
where $\bar{w}(\theta)$ is the mean $w(\theta)$ for the $N=26$ measurements
and $w_i(\theta)$ is the measurement of the galaxy auto-correlation excluding
the $i$th sub-sample.  This same approach will be applied to the data to 
calculate the covariance matrix for all scales in \S\ref{sec:jack-errors}.

\begin{figure}
\plottwo{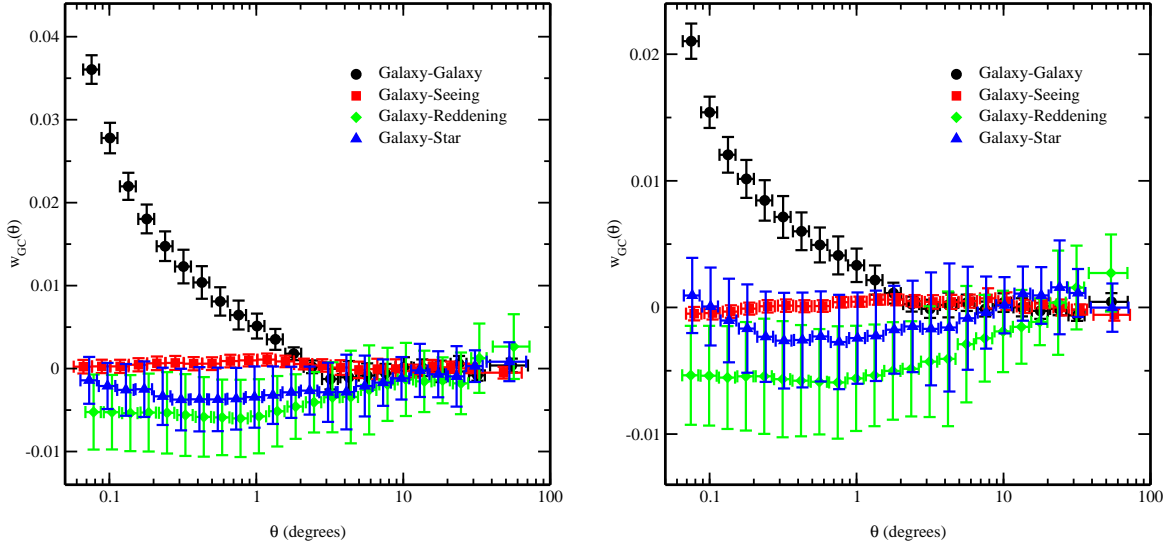}{f20.eps}
\caption{Large-angle cross-correlation of galaxy density and seeing, reddening
and stellar density for the magnitude bins $20 \le r^* \le 21$ (left) and 
$21 \le r^* \le 22$ (right).  The anti-correlation for the reddening and
stellar density is due to variations on the scale of the observed area. 
Since this affects the zero-point for the over-densities, the anti-correlation
occurs on all scales.\label{fig:large-scale}}
\end{figure}

	Figure~\ref{fig:large-scale} shows the cross-correlation between the
galaxy density and seeing, reddening and stellar density using this method for
the faintest two magnitude bins.  As with the results in 
Figure~\ref{fig:all_corr_small}, the galaxy-seeing cross-correlation is 
consistent with zero on all scales for both magnitude bins.  The 
galaxy-reddening and galaxy-star cross-correlations, however, differ 
significantly from the small scales.  This can be understood readily by 
recognizing that the variation in galactic latitude over the course of the 
observing area leads to large-scale variations in the reddening and stellar
density while the seeing variation is basically a small-angle phenomena.  
(Since we calculated the expected values for the contaminants independently 
for each sub-sample in Equation~\ref{eq:contam}, these large-scale variations 
would not factor into those measurements.)  As a result, we see a rather flat 
cross-correlation in the large-angle measurements consistent with zero at the
$1.5\sigma$ level.  The effect of this cross-correlation is the uniform 
inflation of the galaxy-galaxy auto-correlation when calculated on large 
scales, similar to the integral constraint discussed in \S\ref{sec:integral}.

\section{Magnitude System}

	For the actual angular clustering measurements as well as the 
cross-correlation tests in \S\ref{sec:cross}, the magnitude cuts have been 
made using the model magnitudes described in \S\ref{sec:sep_method}.  This 
choice contrasts with a number of other SDSS papers on the EDR which use 
Petrosian magnitudes (Petrosian, 1976).  Our decision to use model magnitudes 
for the angular clustering is based on two points.  First, while Petrosian 
magnitudes are appropriate for relatively bright galaxies, they are not 
intended for stars.  At model magnitudes brighter than $r^* = 19$, stars are 
roughly 0.05 magnitudes fainter in Petrosian magnitudes than in PSF magnitudes 
(there is, of course, no significant disagreement between model magnitudes and 
PSF magnitudes in this range).  At fainter magnitudes, this disagreement blurs 
into considerable scatter between the two magnitude systems, with 
disagreements as large as half a magnitude fainter and 0.1 magnitudes 
brighter. Clearly, since we do not strictly separate between galaxies and 
stars, Petrosian magnitudes are not a proper tool for separating our sample 
into different magnitude cuts.  Additionally, for galaxies, there is a general 
disagreement between the model and Petrosian magnitudes, with the Petrosian 
magnitude for a given object fainter by roughly 0.15 magnitudes, but with a 
much larger scatter than for the stars.  (these large errors are not due
to failures in calculation, but an artifact of the occasionally very large
apertures dictated by the Petrosian method).  

As a result of these differences, applying the same numerical magnitude cuts 
using Petrosian and model magnitudes results in a slightly larger amplitude 
for $w(\theta)$ from the Petrosian sample, relative to the model magnitude 
selection.  For the brightest three magnitude slices, the measurements are 
just within the $1\sigma$ errors on the smallest scales, while for the 
faintest bin, they are only within the $2\sigma$ errorbars.  The 
cross-correlations with seeing and reddening are consistent with zero on all 
angular scales within $1\sigma$ errors.  The galaxy-star cross-correlation is
roughly a third stronger for the faintest magnitude bin using the Petrosian 
magnitude cut, but still considerably smaller than the galaxy auto-correlation.
Switching to Petrosian magnitudes has no effect on the cross-correlation with
sky brightness.  On whole, similar cuts on seeing and reddening would appear
to suffice for galaxies selected using Petrosian magnitudes, but, again, these
magnitudes may not be appropriate for faint objects, particularly those which 
are not easily separated into galaxy and star types.

\section{Masks} \label{sec:mask}

	In addition to making the cuts on seeing and reddening described 
in \S\ref{sec:see} and \S\ref{sec:red}, we also mask out all of the stars 
in the field that have saturated centers, including a rectangular area around 
the star large enough to encompass any diffraction spikes.  Similarly, we 
mask out two thin regions ($\sim 15''$ wide) running the length of the data 
set where the data processing pipeline flagged nearly all of the objects as 
saturated due to a bad CCD column.  As described later in \S\ref{sec:core}, 
poor telescope collimation caused the PSFs in scanlines 1 and 6 to be 
noticeably worse than those in the central scanlines (this problem has since 
been corrected so will not affect any subsequent data).  To analyze the 
possibility of a difference between the central and outer scanlines, we also 
consider a more restrictive mask that excludes scanlines one and six from 
each strip.

\begin{figure}
\plottwo{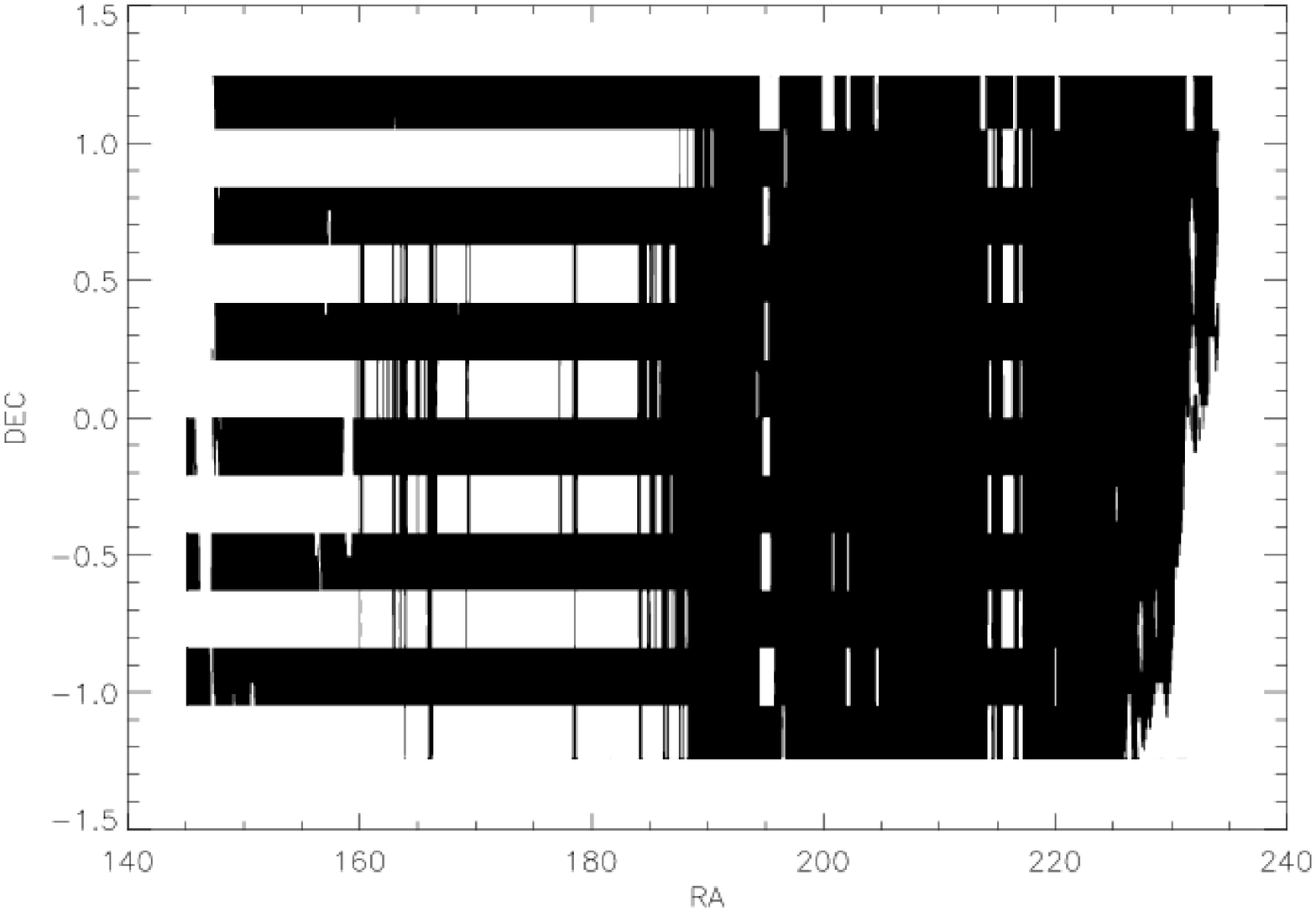}{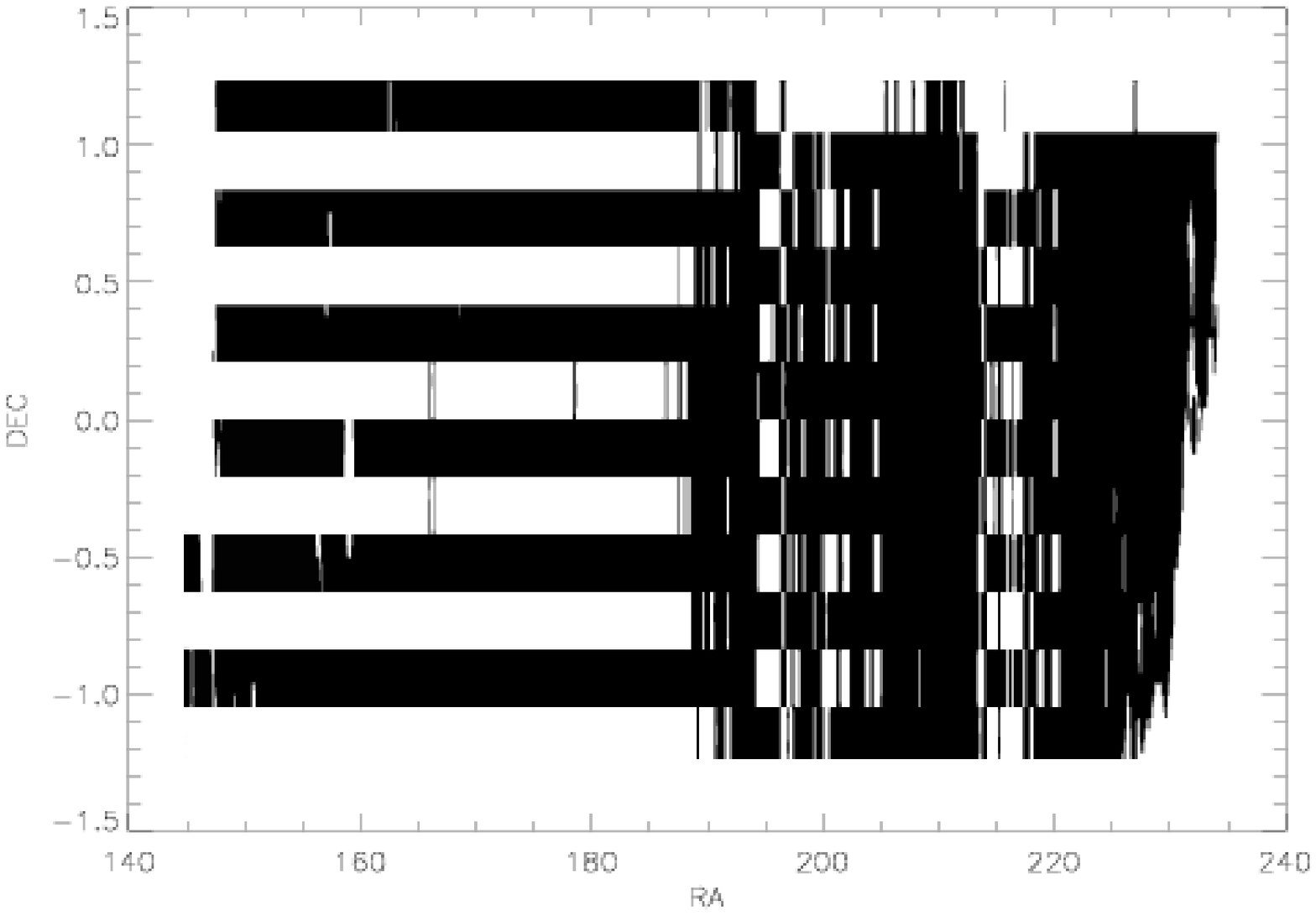}
\caption{The left panel shows the survey area in the limit that the seeing 
was better than $1''.75$ and reddening better than 0.2 magnitudes, as 
appropriate for objects brighter than 21 in $r^{\prime}$.  The right shows 
the same, but with the requirement that the seeing was better than $1''.6$.
This mask was applied to the faintest magnitude bin ($21 \le r^* \le 22$).
\label{fig:bright_mask}}
\end{figure}

Masking out the regions of substandard seeing makes the largest cut
in our data (Figure~\ref{fig:bin_area}), taking out approximately 26\% 
(24\%) of the total (central) area for the bright mask (seeing better than 
$1''.75$); the faint mask (seeing better than $1''.6$) removes 35\% (31\%).  
Imposing our cut on reddening claims roughly another 2\%.  Finally, the area 
lost to bright stars is .05\% of the total area and the two saturated CCD 
columns mask another 0.35\%.  After combining these masks into one unit and 
eliminating overlaps among the different masks, the total area lost to masks 
is 28.5\% (26.8\%) of the total (central) area in the bright mask and 37.9\% 
(33.7\%) for the faint mask.  The effect of the masks, excluding the bright 
star masks, are shown in Figure~\ref{fig:bright_mask}.

\begin{figure}
\plottwo{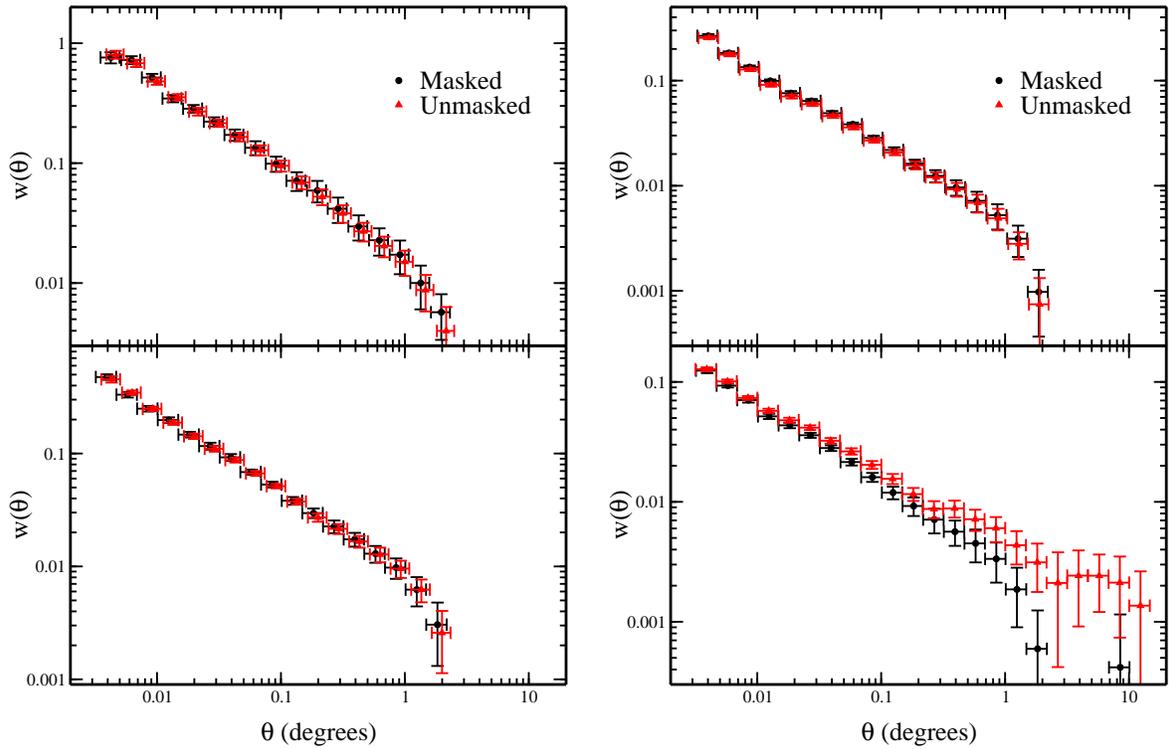}{f24.eps}
\caption{Comparisons of the masked and unmasked measurements of $w(\theta)$
for the four magnitude bins.  The left panel shows the $18 \le r^* \le 19$
(upper) and $19 \le r^* \le 20$ (lower) bins and the right panel shows the
$20 \le r^* \le 21$ (upper) and $21 \le r^* \le 22$ (lower) bins.
\label{fig:masked_unmasked}}
\end{figure}

	Figure~\ref{fig:masked_unmasked} shows a comparison of the masked 
and unmasked $w(\theta)$ measurements for the four magnitude bins.  In 
general, the effect of the mask on the auto-correlation is quite minimal.  
This does not hold true, however, for the faintest magnitude bin, where 
significant departures from the expected power-law shape can be seen, 
coinciding with our expectations of variations on the scanline-scale due to 
cross-correlation with seeing.

	\section{2-D Auto-Correlations}\label{sec:2dcorr}

	For a further check on the performance of the mask and the data, we 
can break our auto-correlations into two dimensions, splitting the angular 
separation between a given pair of objects into its components along the 
scanline and orthogonal to the scanline ($\alpha$ and $\delta$ in our case, 
since our area parallels the Celestial equator).  If we have variations in 
density correlated with the interlaced scanlines, then we should see striping 
along the scan direction. 

\begin{figure}
\plotone{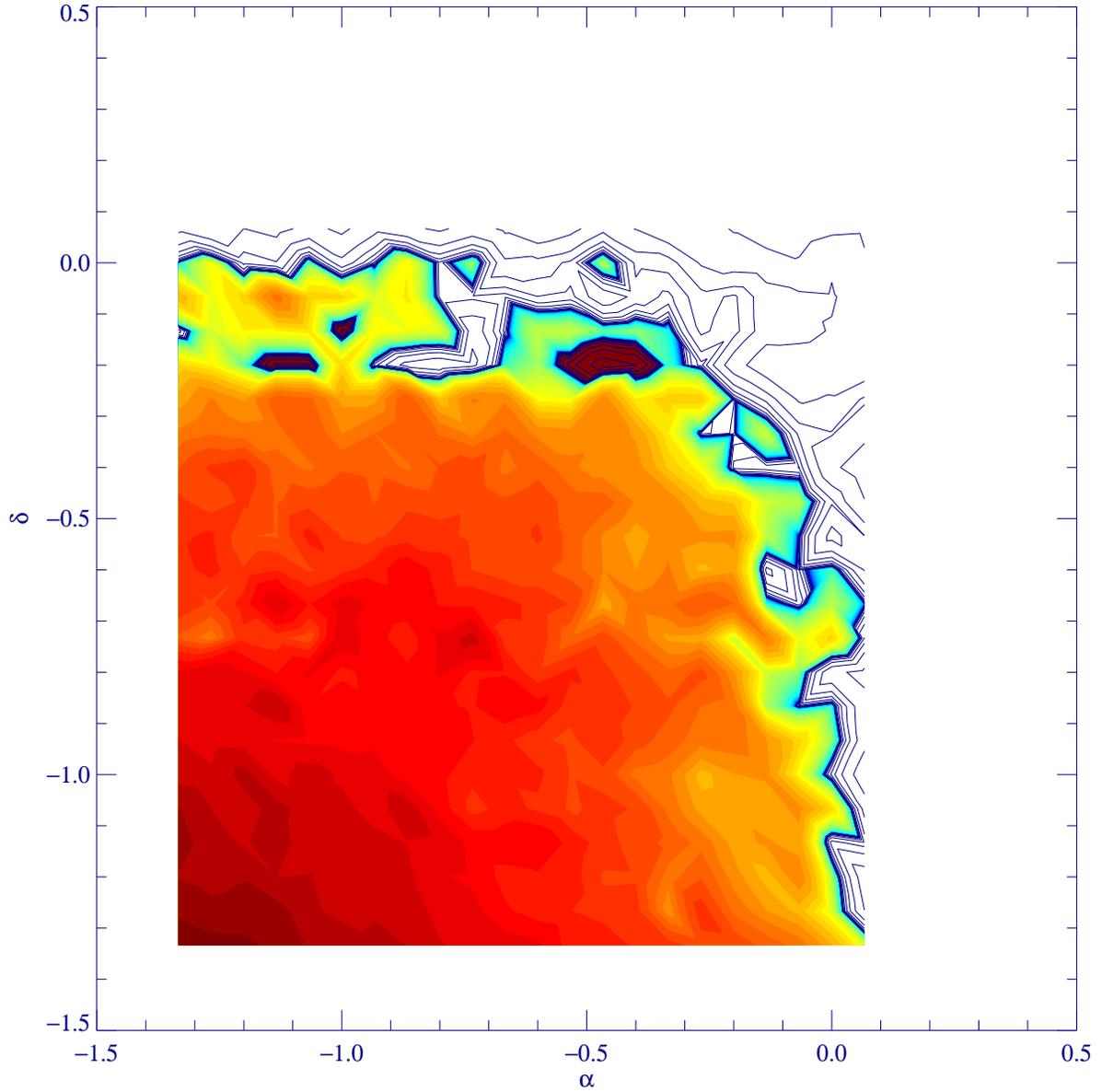}
\caption{Galaxy auto-correlations (positive contours are filled and negative 
are in wire-frame) for the $21 \le r^* \le 22$ magnitude bin with the angular 
separation broken into its component parts along the $\alpha$ and $\delta$ 
axes.  All four magnitude bins show good symmetry in the scanwise and 
orthogonal directions, indicating sufficient masking of bad regions.  X and Y 
axes are the logarithms of the angular bin in degrees in the $\alpha$ and 
$\delta$ directions, respectively. \label{fig:2dcorr}}
\end{figure}

\begin{figure}
\plotone{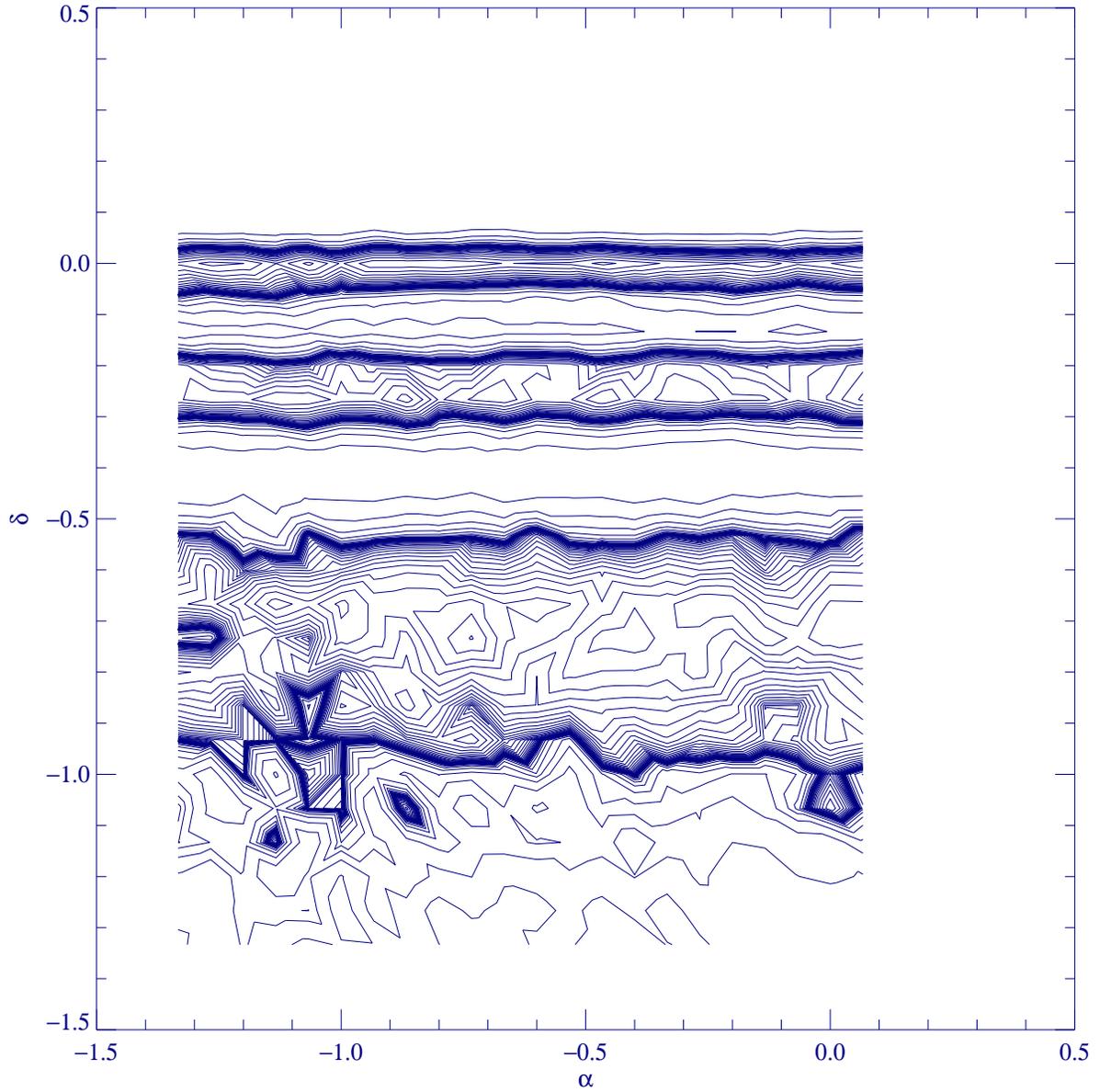}
\caption{Residual difference between the masked and unmasked 2-D galaxy 
auto-correlations for $21 \le r^* \le 22$.  The brighter three bins have 
minimal structure along the scanlines but the faintest bin shows 
significant banding on the scale of the scanlines, a clear sign that the 
mask is necessary to avoid contamination.\label{fig:2dcorr_res}}
\end{figure}

As shown in Figure~\ref{fig:2dcorr}, we recover a radially symmetric 
auto-correlation for the faintest magnitude bin (as well as the three brighter
bins), confirming that our sample is uniform over the transitions between 
scanlines; the slightly rectangular appearance in the plot contours is an 
artifact of the angular binning.  We can also use this measurement to 
demonstrate how the mask improves the auto-correlations.  In 
Figure~\ref{fig:2dcorr_res}, we show the results of subtracting the 
auto-correlation for the faintest bin calculated without the mask from the 
plot shown in Figure~\ref{fig:2dcorr}.  For the three brightest magnitude 
bins, there is no obvious striping due to variations in the galaxy density 
between scanlines, but we do see a rather modest uniform offset, which is 
perhaps due to a decrease in stellar contamination for the masked sample.  
This is not true for the faintest magnitude slice, however, which shows 
exactly the striping in the unmasked measurement we predicted for variations 
in the galaxy density between strips.  The mask removes this behavior to the 
limit of our ability to measure it with this test.  The same exercise can be 
repeated for the stellar population with similar results, albeit with less 
visually impressive results as there is no inherent signal due to clustering.

\section{Camera Column Variations} \label{sec:internal}

	The previous sections dealt with errors that were the result of
external effects on the galaxy data set and we were able to set limits on the
associated contaminants that would produce a uniform galaxy catalog across the 
area of the survey to the best of our ability to measure.  In addition, 
however, uncorrected variations in sensitivity across the imaging camera could 
lead to false correlations. We test for such effects in this section. 

	\subsection{Intra-Column Variations} \label{sec:intra}

	The first check is that the CCDs that create each of the camera 
columns has a uniform depth of field.  Variations in PSF due to telescope 
collimation errors or the like across a given detector could, if not taken 
into account, lead to artificial density variations in object densities and 
classifications.  Improvements in the data processing pipeline have reduced 
the effect of this on our star/galaxy classification and photometry below our 
ability to detect it.

To demonstrate this, we use a finer pixelized version of the data described in 
\S\ref{sec:cross} (cells with sides $\sim 0.01^\circ$ long instead of 
$\sim 0.04^\circ$, using the masks described in \S\ref{sec:mask}) and 
compare the fractional over-density for a given cell $i$, $\delta_i$ 
as calculated in Equation~\ref{eq:delta_def}, in each column to a series of 
over-density gradients ($\Delta_i$) across the chip in the $\delta$ direction. 
We choose stellar density rather than galaxy density as it should be free of 
actual clustering due to large scale structure.  The over-density gradients 
are constructed as sines and cosines:
\begin{equation}
\Delta_{i,n|m} = \left \{ 
\begin{array}{c}  
A_{n} \cos \left (\frac{2\pi n}{\delta_L} \bar{x}_i \right ) \\
B_{m} \sin \left (\frac{2\pi m}{\delta_L} \bar{x}_i \right ) 
\end{array} \right \}
\end{equation}
where $\bar{x}_i$ is the mean declination for cell $i$, $\delta_L$ is the 
width of the column ($\sim 2.52^\circ$), $A_{n}$ and $B_{m}$ are chosen 
such that 
\begin{eqnarray}
\sum_i \Delta_{i,n|m} &=& 0 \\
\sum_i \Delta_{i,n|m} \Delta_{i,n|m} &=& 1,
\end{eqnarray}
and $A_{n} = 0$ for $m > 0$ and vice-versa.  We define
\begin{equation} \label{eq:gradient}
C_{n|m} \equiv \frac{1}{N} \sum_i \delta_i \Delta_{i,n|m},
\end{equation}
where $N$ is the number of pixels and $C_{n|m} = 0$ provided that there is no 
correlation between the two over-densities.  This is effectively decomposing 
the projection of the stellar over-densities along the scan direction into 
Fourier cosine and sine modes.

Since we have 216 cells in the declination direction, we can let $n$ and $m$ 
vary from 0 to 108 and still resolve the variations in $\Delta_{i,n|m}$.  
Calculating the sum in Equation~\ref{eq:gradient}, we should detect any 
fundamental variations orthogonal to the scanlines.  We find, however, that 
the whole of the data satisfies equation~\ref{eq:gradient} to within 
$2\sigma$ of the Poisson errors for all 216 gradients in all four magnitude 
bins.  

We can also repeat this exercise for each scanline independently.  Here, our 
maximum value for $n$ and $m$ is reduced to $9$ and 
$\delta_{L^*} = \delta_L/12$.  Once again, we do not observe any 
statistically significant non-zero elements of $C_{n|m}$ for any of the 
scanlines in either column, indicating that we are not producing correlations 
due to differential response in the camera columns.

\subsection{Cross-Column Correlations}

\begin{figure}
\plottwo{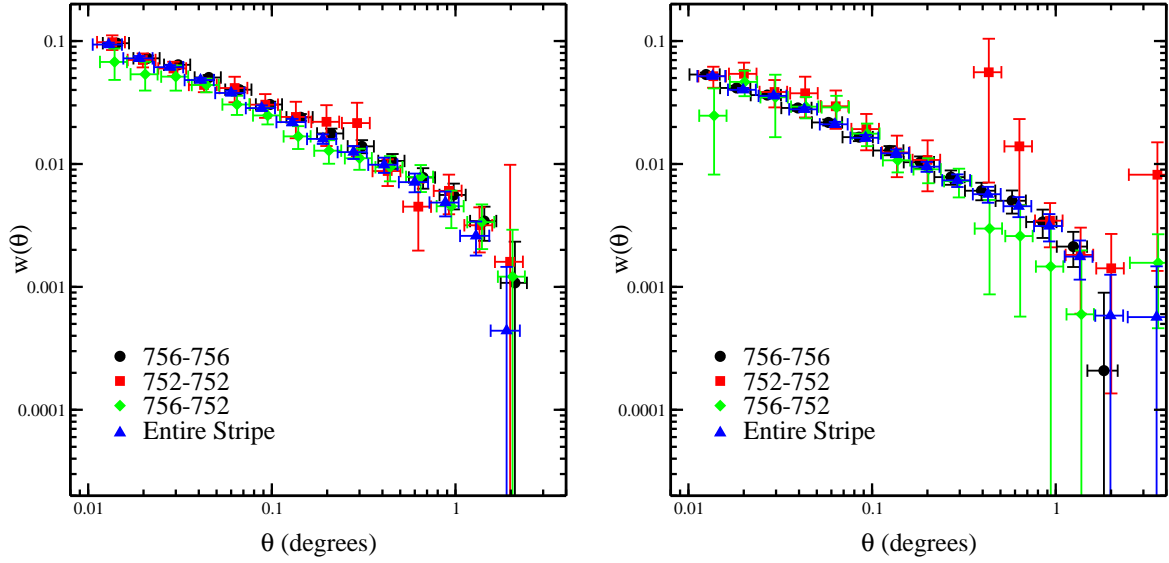}{f28.eps}
\caption{Galaxy auto-correlations for each of the runs in the stripe and the 
cross-correlation between the runs as compared to the auto-correlation for
the whole stripe for the $20 \le r^* \le 21$ (left) and 
$21 \le r^* \le 22$ (right)\label{fig:cross-corr}}
\end{figure}

	As a further check on column-to-column variations, we calculated the
galaxy auto-correlation using the sub-sample method within each of our runs 
independently of the other, as well as the cross-correlation between the 
galaxies in each of the runs.  If our calibrations and sensitivities are 
consistent from run to run, each of these should be consistent with the 
measurement we make from the entirety of the stripe.  The results of this 
measurement using the pixelized data set from \S\ref{sec:intra} and the 
sub-sampling error method from \S\ref{sec:cross} for the two faintest 
magnitude bins are in Figure~\ref{fig:cross-corr}.  In general, they confirm 
that our system is behaving as we would hope.  There is some spurious signal 
from the auto-correlation in Run 752 around the scale size of a scanline, but 
that is likely to be due to the fact that most of that run is masked for the 
first half of the range in right ascension (see Figure~\ref{fig:bright_mask}).

\subsection{Central Scanlines vs. All Scanlines} \label{sec:core}

	The SDSS camera provides an extremely flat depth of field, so as to 
make the observation of objects as they drift across the field as uniform as 
possible.  However, as mentioned above, at the time that these data were 
taken, the telescope was not properly collimated (this has since been 
corrected).  The result was that the PSFs in the outer two scanlines of each 
run were noticeably poorer than the central ones, particularly scanline 6 in 
each run.  This, in turn, made calculation of the photometric calibration more 
difficult in those scanlines.  

As mentioned previously, the photometric calibration is generally done by 
setting photometric zeropoints based upon secondary stars observed by the PT
in the region imaged by the main camera.  The uniformity of this calibration
can be checked by comparing the magnitudes of objects in the $\sim 2'$ overlap 
region between adjacent scanlines.  In $r^*$, the median photometric 
zeropoint offset, as determined from photometry of stars, is less than 2\% 
for all scanlines of data except for scanline 6, which shows deviations of 
up to 5\%.  For galaxies with $r^* < 19$, we find that the rms difference 
between model magnitudes in the two runs in the overlap is typically 0.04 mag 
in $r^*$, about 30\% larger than the nominal photometric errors would 
imply.  

Clearly, this sort of calibration error could potentially lead to a change in 
the galaxy density for a given magnitude range in those camera columns 
compared to the central scanlines, although such a variation is not apparent 
from the galaxy density plots in Figure~\ref{fig:dec_density}.  We tested 
this possibility by excluding the outer scanlines and re-calculating the 
cross-correlations and auto-correlations from \S\ref{sec:cross}.  In general, 
the central scanlines were somewhat less sensitive to systematic contaminants 
than the whole of the data, but it appears that adding the outer scanlines 
does not effect the resulting auto-correlation.

\section{Limber Scaling Tests}

To test the consistency of the $w(\theta)$ measurements and check whether the 
variations in $w(\theta)$ are due to intrinsic clustering effects rather than 
systematic errors, a magnitude scaling test using the relativistic form of 
Limber's equation was applied to the results.  Since $w(\theta)$ is given by 
the two-dimensional projection of the spatial clustering function $\xi(r)$, 
the $w(\theta)$ measurements will be scaled according to the depth of the 
survey (Peebles, 1980). 

With a model for the redshift distribution $dn/dz$, we can
scale the measurements of $w(\theta)$ in disjoint magnitude slices to the 
same depth. This method is essentially identical to the one employed in the 
APM survey (Maddox et al. 1996), described there in greater detail. We 
assume the same two-slope form of $\xi(r)$, with slopes of $\gamma=1.7$ at 
small angles and $\gamma=3.0$ at large angles, and ignore effects of 
evolution in $\xi(r)$.

As with the simulated catalogs discussed in \S\ref{sec:mock} below, the
$dn/dz$ selection function was based on the results of the CNOC2 survey (Lin 
et al., 1999), which used a parametrized model for the evolution of the galaxy 
luminosity function. It assumed a standard Schechter function with no 
evolution in $\alpha$ and fit a linear evolution model for $M^{\star}$ to 
$M^{\star}(z=0.3)$, which was the mean redshift for CNOC2 galaxies. 
Calculations were performed for two different cosmologies: a critical-density 
universe ($\Omega_{m}=1, \Omega_{\Lambda}=0$) and an under-dense model 
dominated by a cosmological constant ($\Omega_{m}=0.3$, 
$\Omega_{\Lambda}=0.7$). 

Because the commissioning data form a narrow stripe, it does not allow us to 
accurately calculate $w(\theta)$ at angles larger than $\sim 1^{\circ}$, so 
we cannot observe the expected break in the power law. The break from power 
law form should occur at larger angles for fainter magnitudes, which is 
accounted for in the scaling tests, but we cannot verify this using the 
commissioning data.

\begin{figure}
\plottwo{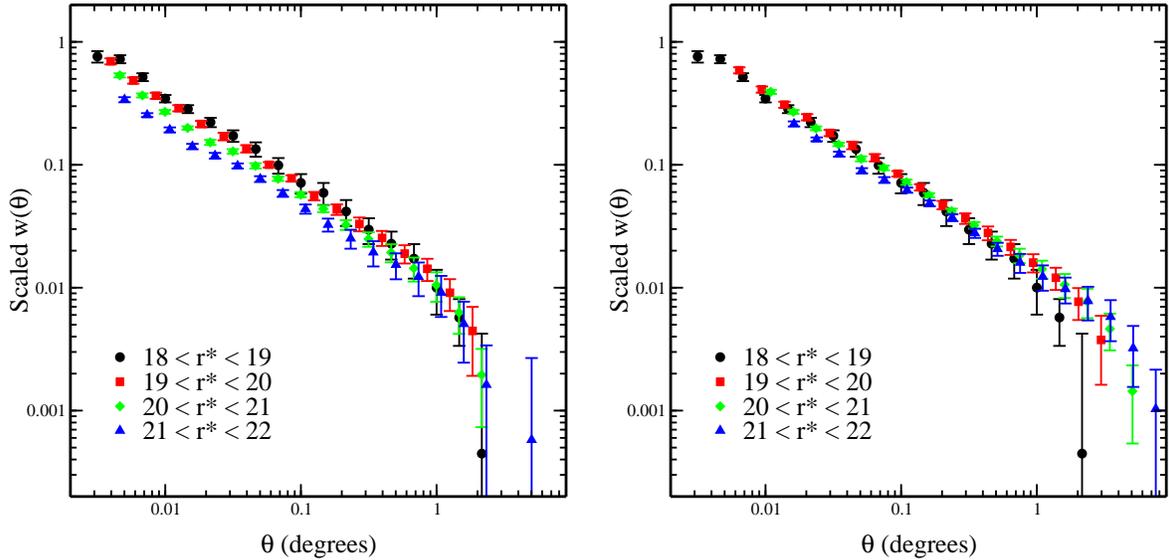}{f30.eps}
\caption{Limber scaling tests for the four magnitude bins assuming a flat,
matter-dominated cosmology (left panel) and flat, $\Lambda$-dominated 
cosmology (right).  In both cases, the measurements in the fainter bins have 
been scaled to the brightest magnitude bin.\label{fig:limber_test}}
\end{figure}

As seen in Figure~\ref{fig:limber_test}, the critical-density calculations 
show good agreement for the brighter magnitude bins, with progressive 
worsening for the fainter magnitudes. This does not contradict the APM 
measurements (Maddox, 1996), since the APM survey had a magnitude limit of 
$b_{J}=20.5$ and the discrepancy only becomes significant for the faintest 
magnitude bins, due to the strongly differing size of the volume element at 
large ($z>~0.5$) redshifts between the cosmological models examined.

The lambda-dominated tests show much better agreement across all magnitude 
ranges. The fact that the scaling tests should support an under-dense, 
lambda-dominated universe is not surprising, since measurements of $dn/dz$ of 
faint galaxies (Fukugita et al. 1990) have been known to be incompatible with 
a critical-density universe for some time.

\section{Deblending Tests}

	As our final source of systematic errors, we consider the possibility 
of errors in deblending nearby pairs of objects during data processing.  Since 
the data pipeline is automated, the software is required to make decisions as 
to what is a single object and which are close projections of two 
distinct objects on the sky.  

The expected primary source of deblending problems for the magnitude range 
we are considering is the combination of two distinction objects which cannot 
be successfully separated.  This can happen under a number of circumstances, 
most notably seeing variations that blur the object images to the point where 
even pairs of morphologically simple galaxies cannot be clearly separated.  Of 
course, since we are interested in extended objects, we are more sensitive to 
these errors than we would be if we were only interested in stellar objects.  
We have the additional concern that the objects causing this sort of error 
need not be in our magnitude cut, but rather could be stars or galaxies in 
different magnitude bins.  

The primary effect of this error is on the smallest angular scales, where we 
expect to see a suppression of the number of pairs of objects, as multiple 
objects in high density regions are interpreted as single objects.  This effect
can then propagate to larger angles due to the expected high correlation 
between angular bins.  Likewise, since we will have mis-counted the number of 
objects, our estimation of the number density of objects on the sky could be 
significantly skewed, resulting in a suppression of the auto-correlation 
signal similar to an integral constraint (as discussed in 
\S\ref{sec:integral}) on all angular scales.  

	\subsection{Input Catalogs}

Since we do not know a priori what the nature of the deblending errors would
be in our data, we need a training set of perfectly resolved data that could 
then be manipulated to simulate various deblending failures.  For this 
purpose, we use mock catalogs (described in section \S\ref{sec:mock}) that 
generated for the measurement of the $w(\theta)$ covariance matrix.  
Since we expect that the rate of deblending error is highly 
dependent on the density of objects on the sky, we concentrate on the mock 
catalogs simulating the faintest magnitude bin.  The other three magnitude 
bins as well as a randomly distributed set of points with projected density 
approximately equal to the stars were used to simulate foreground objects 
that might also cause deblending problems.

	\subsection{Small-Angle Supression Test}

	To measure the suppression of the small-angle signal due to deblending
errors, we concentrated solely on the possible failure of the deblender to
separate close pairs of objects.  To simulate the efficacy of the deblender, 
we calculated the separation of all objects within the main sample and the 
separation on the sky of those objects and objects in the foreground sample.
This separation ($\Delta\theta$) was used to generate a probability of being 
successfully separated using a sigmoid function of the form
\begin{equation}
P(\Delta\theta) = \left ( 1 + e^{-(\Delta\theta-\sigma_c)/\sigma_s} 
\right)^{-1},
\end{equation} 
where $\sigma_c$ represents the separation at which half the objects are 
successfully deblended and $\sigma_s$ controls the slope of the likelihood.
Applying this treatment to the data resulted in a sample in which the 
likelihood of close objects being successfully deblended decreases according 
to their relative separations.   We let $\sigma_c$ vary from 0'' to 5'' and 
$\sigma_s$ from 0 to 1.25 ($\sigma_s = 0$ indicating a step-function at 
$\Delta\theta = \sigma_c$), calculating $w(\theta)$ for each 
combination and comparing the result to the observed suppression in the data:
\begin{equation}
\delta w(\theta) \equiv \frac{w_T(\theta) - w_D(\theta)}{w_T(\theta)},
\end{equation}
where $w_T(\theta)$ is the value expected from the template and $w_D(\theta)$ 
is the measurement with deblending errors.  For the actual data, we made the 
assumption that the ``true'' $w(\theta)$ is well-described as a power law; 
Connolly et al. (2001) gives the parameters for this fit.  Treating this power
law as the $w_T(\theta)$ for the real data, we found the combination of 
$\sigma_c$ and $\sigma_s$ that produced residuals similar to the residuals in
the data.  The best fitting values for $\sigma_c$ and $\sigma_s$ are 3'' and 
0.5, suggesting that we cannot trust the deblender to function on this data 
at better than 95\% efficiency for angular scales smaller than $\sim 6''$.  
In the mock catalogs, this level of deblending errors reduced the number of 
galaxies by 2.6\%, small enough not to have an apparent effect on the overall 
integral constraint (see \S\ref{sec:integral}).

\begin{figure}
\plotone{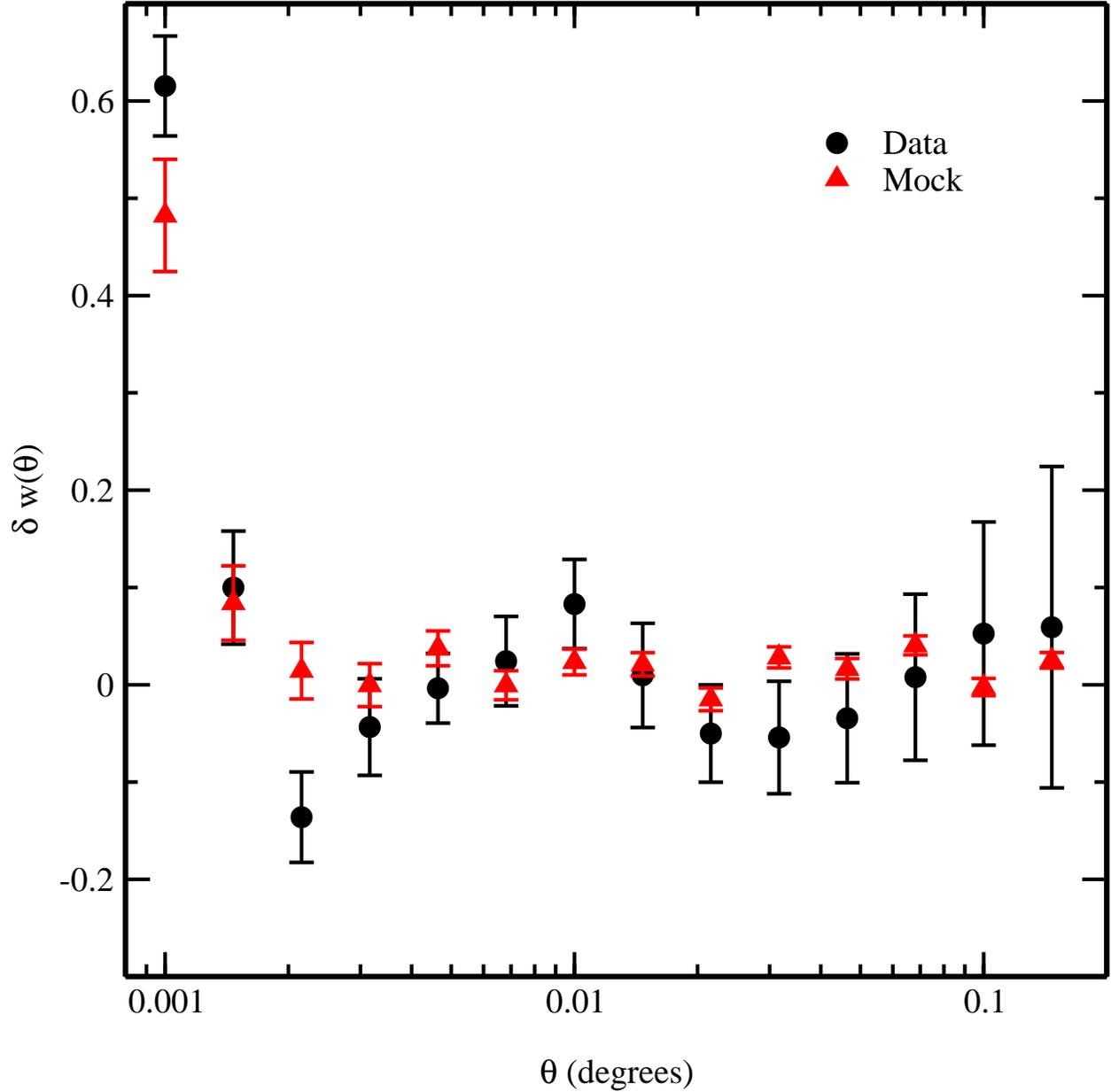}
\caption{The $\delta w(\theta)$ residuals for the best fitting model of the 
deblending errors.  In both the data and the simulation, aliasing of power 
and non-zero covariances between angular bins lead to variations which are, 
however, consistent with zero for angular scales $> 6''$.  
\label{fig:deblend_err}}
\end{figure}

Although the data on angular scales larger than 6'' is consistent 
with a power-law and the mock catalog with the simulated deblending errors is 
consistent with the template measurement, the residuals plotted in 
Figure~\ref{fig:deblend_err} suggest that the lower angular limit on 
deblender efficiency generates periodic variations in the data.  The suspected
source of these variations is the aliasing of power into the third
angular bin from the first two and the large covariance between the angular 
bins.  These variations are consistent with zero for this measurement, but 
improvements in the errors due to a large observing area would likely make 
them discordant.  This suggests that future SDSS measurements of $w(\theta)$ 
will need to take deblender effects into account to avoid misleading signals 
on all scales.  Additionally, one might mitigate these effects by looking at
magnitude-weighted angular correlations.

\section{Estimators \& Biases} \label{sec:estimators}

Having checked both the internal and external sources of systematic errors
which might apply to any angular clustering or photometric work, as well as 
developing a mask and angular limit to avoid regions where we would have 
significant systematic errors, we are now ready to address more specific
issues related to the measurement of $w(\theta)$ and the associated covariance 
matrices.

\subsection{Estimators}

We now turn our attention to the estimators for $w(\theta)$.  There are a 
number of estimators in the literature (Peebles (1973), Sharp (1979), 
Hewett (1982), Landy \& Szalay (1993), and Hamilton (1993)), but we have 
generally relied on two:
\begin{equation}
\hat{w}(\theta) = \frac{DD - 2 \langle DR \rangle + 
\langle RR \rangle}{\langle RR \rangle},
\label{eq:particle}
\end{equation}
where $DD$ is the number of galaxy pairs in a given angular
bin, $\langle RR \rangle$ is the expected number of random pairs for a random
catalog of similar density and geometry and $\langle DR \rangle$ is the 
expected number of cross-population pairs; and
\begin{equation}
\hat{w}(\theta_\alpha) = \frac{\sum_{i,j} \delta_i \delta_j U_i U_j 
\Theta^\alpha_{ij}} {\sum_{i^\prime,j^\prime} U_i U_j 
\Theta^\alpha_{i^\prime,j^\prime}},
\label{eq:pixel}
\end{equation}
where $\delta_i$ is the fractional overdensity in cell $i$ (as given in 
Equation~\ref{eq:delta_def}), $U_i$ is the fraction of cell $i$ that is 
unmasked, and $\Theta^\alpha_{i,j}$ is 1 if cells $i$ and $j$ are separated by 
a distance in angular bin $\alpha$ and zero otherwise.  This is similar, of 
course, to the estimator in \S\ref{sec:cross}, but with a cell size determined 
by the desired angular resolution and terms to deal with a mask that is 
independent of the cell size. The estimators in Equations~\ref{eq:particle} 
and \ref{eq:pixel} are identical in the limit of infinitely small cell sizes
(Szapudi \& Szalay (1998)).

In its simplest form, the first estimator has the advantage in that it can, in 
principle, probe all of the angular scales in a fixed amount of time.  
Traditionally the time to perform this calculation goes as $\mathcal{O}(N^2)$.
We can take advantage of the shape of the data area to speed up the
calculation, sorting the objects by $\alpha$ and only considering pairs 
separated by angles less than the largest angular bin.

While this offers an improvement, the large numbers of measurements 
necessary to calculate the covariance matrices (see \S\ref{sec:mock}) 
require a more sophisticated technique.  For this calculation, we take 
advantage of {\it kd-trees} ($k$-dimensional data tree structures 
describing the distribution of the data) to make the calculation run more or 
less linearly with the number of galaxies.   The power of kd-trees for 
pair counting calculations, as developed by Friedman, Bentley and 
Finkel (1977), comes in the quick elimination of large fractions of the 
data, reducing the number of distance calculations for each pair of objects.  
This is accomplished by recursively subdividing the data area into smaller 
nodes (generally by splitting along the widest axis of the data area) until 
sufficient resolution is achieved, one object per node in our case.  A search 
for objects within some radius of a given point in the data area can simply 
trace back up the data tree until the nodes pass out of its accepted radius, 
avoiding most of the data in the process.  In addition, one can calculate 
numerous statistics at each node (count, centroid, covariance, etc.) to create 
an {\it mrkd-tree}, improving the ability of the code to determine whether it 
should progress to the next node to find viable pairs.  To make our 
calculations, we used a version of the {\it mrkd-tree} code (the NPT code) 
developed by Gray \& Moore (2001).  The nature of this code makes it 
extremely fast for scales $\le \sim 0.2^\circ$, but it bogs down for larger 
angles such that we prefer to use the pixelized version of the code for those 
scales. 

\begin{figure}
\plotone{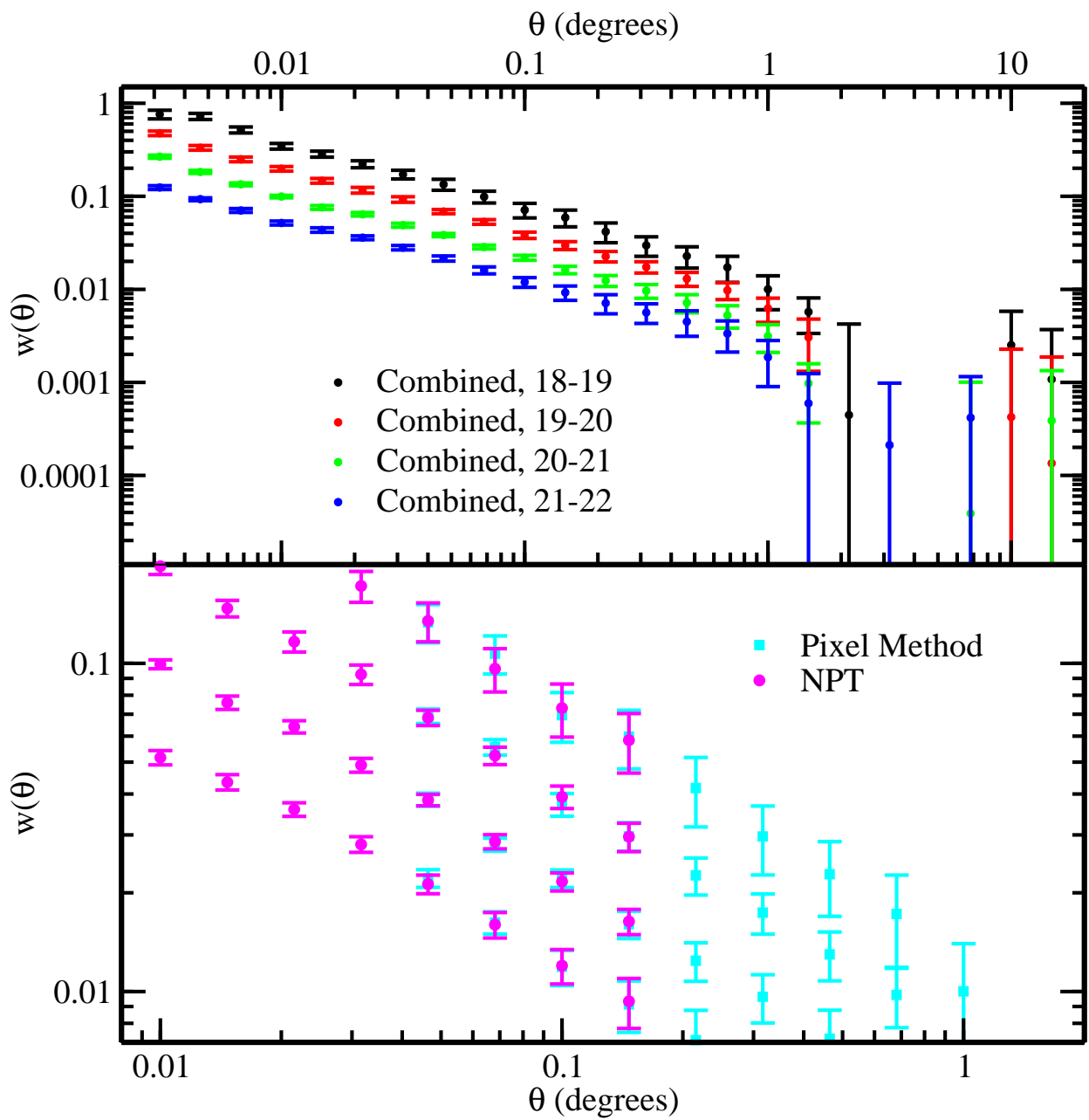}
\caption{The upper panel shows the results of $w(\theta)$ measurements using 
the estimator in Equation~\ref{eq:particle} for angular scales less than 
$0.15^\circ$ and the estimator in Equation~\ref{eq:pixel} for angular scales 
larger than $0.04^\circ$.  The lower panel shows the agreement between the 
two estimators in the overlapping angular bins.\label{fig:data_combine}}
\end{figure}

The virtues of the second estimator are more apparent on large scales, where 
the number of pixels needed is much less than the number of galaxies.  It 
also has the advantage of a more natural generalization to dealing with 
correlations between continuous phenomena (e.g., galaxy density and seeing),
allowing us to use it for the cross-correlations in \S\ref{sec:cross}.  
We have verified that these two methods give comparable measurements of 
$w(\theta)$ over two decades of our angular bins (at 6 bins per decade in 
degrees), although in the full measurement and calculation of the errors we 
restricted the angular overlap region to four bins.  By approximately 
matching the processing time for the two codes, we used the pair counting 
estimator for scales from $0.001^\circ$ to $0.15^\circ$ and the pixelized 
version for scales from $0.04^\circ$ to our upper limit at $14.7^\circ$.
This gives us 26 angular bins, 14 for the pair counting estimator and 18 for 
the pixelized estimator.  Figure~\ref{fig:data_combine} shows the 
results of combining these two estimators to give the full range of angular
measurements, as well as the measurements for each technique in the overlap 
range.  As expected the two methods agree very well and the transition is 
quite smooth. 

       \subsection{Integral Constraint} \label{sec:integral}

        Regardless of the method used, all estimators for $w(\theta)$ are 
plagued by the ``integral constraint''.  Again, there has been considerable 
work done on this problem in the literature (e.g. Peebles 1980, 
Landy \& Szalay 1993, Hamilton 1993, Bernstein 1994, Tegmark et al. 1998, 
Hui \& Gazta\~naga 1999, Szapudi et al. 1999).  The central problem in these 
calculations lies in the fact that the calculation of the mean number density 
($\bar{n}$) for a given cell is not the ``true'' number density, but only an 
estimator thereof ($\hat{\bar{n}}$) based upon a finite number of galaxies 
and cells.  This estimator enters into the estimator of the auto-correlation 
$\hat{w}(\theta)$ in a non-linear fashion and generally tends to suppress our 
estimate of the ``true'' $w(\theta)$.  We will explicitly give the corrections 
only for the pixelized estimator (Equation~\ref{eq:pixel}), but the treatment 
for the particle-based estimator (Equation~\ref{eq:particle}) follows similar 
lines.

                \subsubsection{Bias Correction}

        Since our estimator for $\hat{\bar{n}}$ enters in the calculation of
the over-density, we have to re-define $\hat{\delta}$ as 
\begin{equation}
\hat{\delta_i} \equiv \frac{n_i - \hat{\bar{n}}}{\hat{\bar{n}}} =
(\delta_i - \alpha)(1 - \alpha + \alpha^2 - \ldots),
\label{eq:delta}
\end{equation}
where $\delta_i$ is the true overdensity for pixel $i$ and we have 
parameterized our bias due to using a finite sample with $\alpha$, 
\begin{equation}
\alpha \equiv \frac{1}{N}\sum_i \delta_i,
\label{eq:alpha}
\end{equation}
where $N$ is the number of pixels in the survey area.  Plugging
Equations~\ref{eq:delta} and \ref{eq:alpha} into \ref{eq:pixel}, it
can be shown that 
our estimator for $w(\theta_\beta)$ has an expectation value of
(see e.g. Hui \& Gazta\~naga 1999 for details)
\begin{equation}
\left \langle \hat{w}(\theta_\beta) \right \rangle =
w(\theta_\beta) - \frac{1}{N^2}\sum_{i,j} w(\theta_{i,j}) - 
\frac{2}{N^2} \sum_{i,j,k} w_3(\theta_{i,j,k}) W_{ij,\beta} + 
w(\theta_\beta) \frac{3}{N^2} \sum_{i,j} w(\theta_{i,j}).
\label{icbias}
\end{equation}
where shot noise is excluded (which can be shown to contribute
negligibly to the integral constraint bias), and 
where $W_{ij,\beta}$ is our window function,
\begin{equation}
W_{ij,\beta} = \frac{\Theta^\beta_{ij} U_i U_j}
{\sum_{i^\prime,j^\prime} \Theta^\beta_{i^\prime,j^\prime} U_i U_j}.
\end{equation}
The above expression for $\left \langle \hat{w}(\theta_\beta) \right \rangle$
keeps all terms that are first order in $N^{-2} \sum_{i,j} w(\theta_{i,j})$
(assuming $w_3$ is related to $w$ in the usual hierarchical fashion),
which can be regarded as the small parameter of our expansion.
Note that this expression does not assume $w (\theta_\beta)$ is itself
small. The second term on the right hand side of Equation 
\ref{icbias} is what Peebles (1980) originally derived for integral constraint
correction. 

To estimate the integral constraint correction, we use the hierarchical
relation to approximate the $w_3(\theta)$ term as
\begin{equation}
- \frac{2}{N^2} \sum_{i,j,k} w_3(\theta_{i,j,k}) W_{ij,\beta} = 
- w(\theta_\beta)\frac{2c_{12}}{N^2} \sum_{i,j} w(\theta_{i,j}),
\end{equation}
where $c_{12} \equiv w_3/w_2$ is the hierarchical amplitude.  Bernardeau 
(1994) gives the value of this parameter as $c_{12} = 68/21 + \gamma/3$, where 
$\gamma$ is the logarithmic slope of the variance and galaxy biasing is 
ignored.  For reasonable values of $\gamma$, this gives $c_{12} \sim 2$.
Gathering terms, this gives us our correction ($\Delta \hat{w}(\theta)$):
\begin{equation}
\Delta \hat{w}(\theta_\beta) \equiv \left \langle \hat{w}(\theta_\beta) 
\right \rangle - w(\theta_\beta) = \left (1 + (2c_{12} - 3)w(\theta_\beta) 
\right ) \frac{1}{N^2} \sum_{i,j} w(\theta_{i,j}),
\label{eq:bias}
\end{equation}
We note that the $w_3(\theta)$ term does not in fact contribute
significantly to the bias on large scales (where the bias
is most important), and so the approximations
made above do not unduly affect our estimate of $\Delta \hat{w}$ .

Note that this expression contains $w(\theta)$ and not the estimator 
$\hat{w}(\theta)$ that we have calculated.  However, since the sums involved 
in Equation~\ref{eq:bias} are suppressed by a factor of $1/N^2$, the total 
amplitude of the correction is quite small compared to the value of 
$\hat{w}(\theta)$ on most scales.  This means that the error that we would 
make by substituting the estimator values for $w(\theta)$ into 
Equation~\ref{eq:bias} should also be quite small.  Even if this is not quite 
the case for all scales, we can note the amplitude of the correction to 
$\hat{w}(\theta)$ and disregard those measurements where the correction is a 
sizable fraction of the original estimator.  This will come into play in the 
next section when we consider the applicability of the various error 
calculations for the estimators.

        \subsubsection{Magnitude of the Bias Correction}

\begin{figure}
\plotone{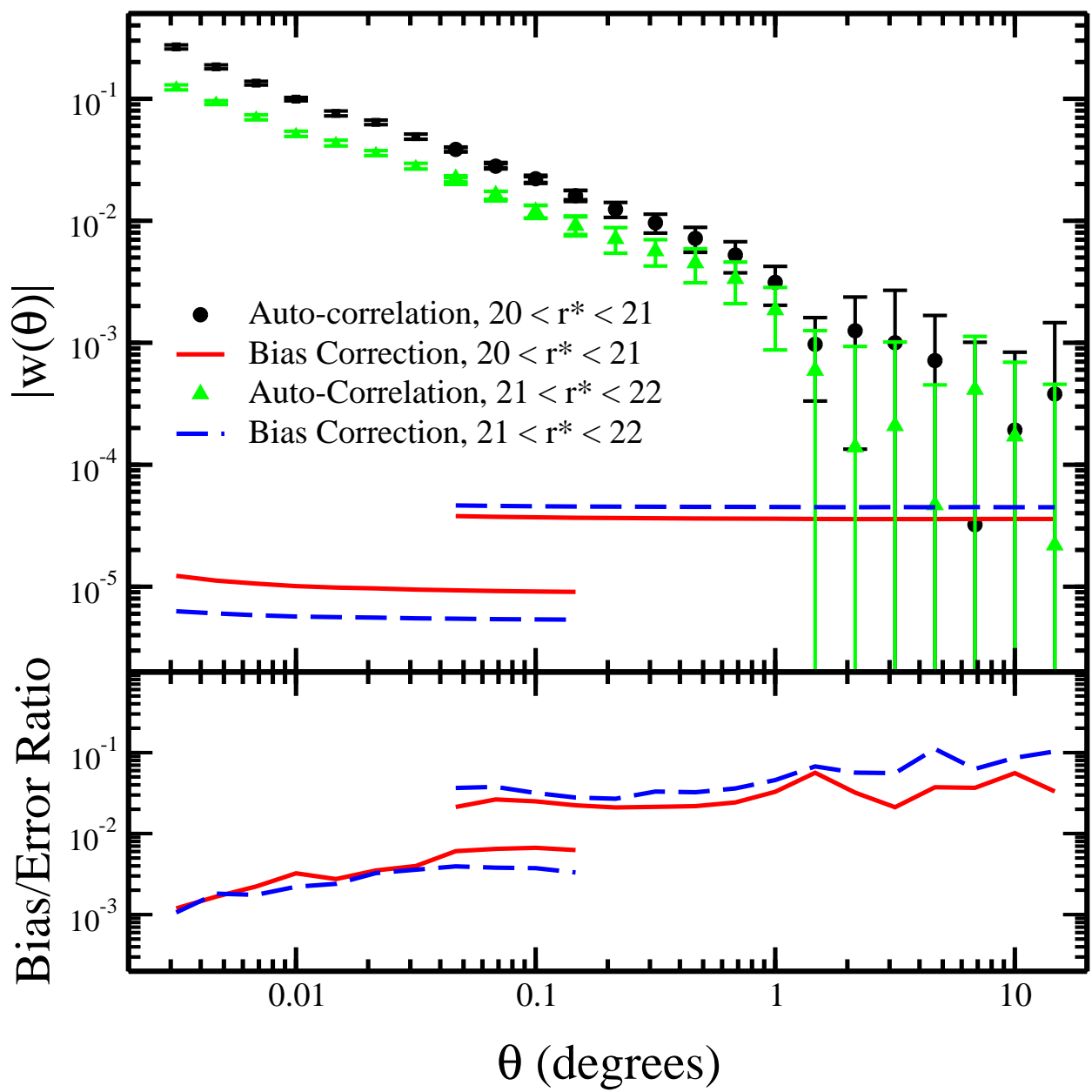}
\caption{Comparison of the galaxy auto-correlation and its errors to the 
integral constraint bias correction suggested by equation~\ref{eq:bias}.  
The different bias correction levels are due to the difference between the
number of pixels and number of objects in the large-angle and small angle
techniques, respectively.  The lower panel gives the ratio of the bias
correction and the error on $w(\theta)$.\label{fig:bias_corr}}
\end{figure}

        In Figure~\ref{fig:bias_corr}, we plot the integral constraint bias 
suggested by equation~\ref{eq:bias} for the faintest two magnitude bins,
comparing them to the auto-correlation and the error on the auto-correlation
as determined using the simulations described in \S\ref{sec:mock}.  The 
difference between the number of objects and pixels used in the estimators 
from  Equations~\ref{eq:particle} and \ref{eq:pixel}, respectively, lead to 
different integral constraint corrections on small and large angular scales.  
In all cases, however, the magnitude of the integral constraint correction 
remains very small relative to the magnitude of $w(\theta)$ suggesting that 
our approximation of $w(\theta)$ by $\hat{w}(\theta)$ is fairly well justified.

\section{Error Calculation and Correlation}

	As with any measurement, the calculation of the auto-correlation is 
only the first step; equally important is the determination of the error 
matrix.  While the Fourier modes in the density field are expected to 
evolve independently (in the linear approximation), a given angular bin will
sample a combination of those modes.  This demands that we calculate the 
correlations between angular bins as well as the standard diagonal elements of 
the covariance matrix if we want to be able to use this measurement in a 
meaningful way.  Of course, we have to find a method of error calculation that 
can be practically and reliably applied over more than three decades in 
angular scales.  Unfortunately, there is no single method of error calculation 
that can do so with the data available alone.  Rather, we use simulations to 
provide us with multiple realizations for error calculation and then check 
those errors against less preferred data-based methods.  Likewise, on large 
scales, we check the simulation covariances against those calculated from
the data under the assumption of Gaussianity.  The next four sub-sections 
outline each of the methods and present examples of the respective covariance
matrices.

	\subsection{Errors from Simulations} \label{sec:mock}

The mock catalogs were generated using a new algorithm 
(Scoccimarro \& Sheth 2001) called PTHalos. PTHalos works in two steps, first 
it generates the large-scale dark matter distribution using second-order 
Lagrangian perturbation theory (2LPT), which reproduces the correct two and
three-point statistics at large scales, and approximates four-point statistics 
and higher (Moutarde et al. 1991; Buchert et al. 1994; Bouchet et al. 1995; 
Scoccimarro 2000) very well. The second step builds up the small-scale 
correlations using the amplitude of the 2LPT density field to determine the 
masses using the algorithm in Sheth \& Lemson (1999) and positions of halo 
centers, and then distributing particles around the halo centers with NFW 
profiles (Navarro, Frenk \& White 1997).  Thus in PTHalos, the large-scale 
correlation function are the result of the perturbative growth of structure, 
whereas the small-scale behavior is due to the internal structure of 
virialized halos. In addition, a ``galaxy'' distribution can be generated in 
PTHalos by specifying how many galaxies populate halos of a given mass. Thus, 
it is possible to design ``galaxy'' distributions which approximate the 
observed statistics; for this purpose, we use a similar grid of models to 
those in Scoccimarro et al. (2001).

The advantage of this method is that it approximates very well the
fully non-linear, and thus non-Gaussian, evolution of structure in a
very small fraction of the time and cost of a full $n$-body
simulation; PTHalos takes about 10 minutes on a DEC Alpha to generate a mock 
catalog for the four magnitude bins used in this paper that would otherwise 
cost several expensive hours of CPU.  Given the uncertainty involved in 
modeling galaxy biasing, the approximate nature of PTHalos distributions is a 
small price to pay in exchange for speed. In particular, the increase in speed 
means we can run many (of the order of a hundred) realizations of the survey 
area, and compute errors and covariance matrices for the clustering statistics
from a Monte Carlo pool. Therefore, our errors automatically take into
account the non-Gaussianity of the galaxy distribution, cosmic
variance, shot noise, and the geometry of the survey.

The catalogs are constructed from parent 2LPT simulations containing
54 million particles that correspond to a $\Lambda$CDM model
($\Omega=0.3$, $\Omega_\Lambda=0.7$; at $z=0$).  The evolution of structure
along the past light-cone is done approximately, by tiling boxes with different
values of the power spectrum normalization $\sigma_8$.  A ``low redshift'' box
of side 600 $h^{-1}{\rm Mpc}$ is used for $z < 0.4$ with 27 million particles 
and $\sigma^{\rm mass}_8 = 0.83$.  Another box of side 1200 $h^{-1}{\rm Mpc}$
and same number of particles with $\sigma^{\rm mass}_8 = 0.66$ is used for 
$0.4<z<0.8$.

Galaxies are assumed to populate dark matter halos
according to the relation (similar to that in Kauffmann et al. 1999,
and Sheth \& Diaferio 2001) 
\begin{equation}
N_{\rm gal} = 0.7 x^{0.8}+y^{0.9},
\label{eq:SD}
\end{equation}
\noindent where $N_{\rm gal}$ is the mean number of galaxies per halo 
of mass $m$. Here $x=m/4\times 10^{12}M_{\sun}/h$ with $x=1$ for 
$m<4\times 10^{12}M_{\sun}/h$, and $y=m/2.5\times 10^{12} M_{\sun}/h$. 
In addition, $N_{\rm gal}=0$ for $m \leq 3\times 10^{11}M_{\sun}/h$.  Such a 
biasing relation between galaxies and mass leads to a large-scale bias 
parameter $b \sim 0.7$.

To generate the galaxy distribution we used the following radial selection 
functions ($dN/dz \propto z^2 \psi(z)$):
\begin{equation}
\psi(z)=\frac{A}{z^{a}} \exp(-(z/z_0)^2) \label{eq:selection_function},
\end{equation}
where $A=6.89, 28.34, 36.42$, $a=1,0.5,0.5$ and $z_0=0.21,0.25,0.35$
for magnitudes bin $18 \le r^* \le 19$, $19 \le r^* \le 20$ and 
$20 \le r^* \le 21$, respectively.  For the faintest magnitude bin, we 
had to use the sum of two selection functions; the first was of the form in 
Equation~\ref{eq:selection_function} with $A=777.35$, $a=0.5$ and
$z_0=0.21$ and the second with the form 
\begin{equation}
\psi(z)=\frac{A}{z^{a}} \exp(-(z/z_0)^3) \label{eq:selection_function2},
\end{equation}
where $A=50.025$, $a=0$ and $z_0=0.545$.  These selection functions are 
based on the CNOC2 luminosity functions (Lin et al. 1999).  The details of 
this extraction and the conversion to SDSS filters are given along with the 
inversion of our $w(\theta)$ measurement in Dodelson et al (2001), which also 
provides an alternate parameterization.

The three-dimensional data is then projected (using different sections
of the simulation box without repeating any structures) into an
angular stripe of 2.5 by 90 degrees, with the same angular coverage to
the actual data from 752-756 runs.  In addition to the comparisons between
the measurements of $w(\theta)$ in the simulated and real data sets described 
below, we have also verified that higher-order moments such as skewness and 
kurtosis match the observed ones to a good approximation 
(Szapudi et al. 2001). 

\begin{figure}
\plottwo{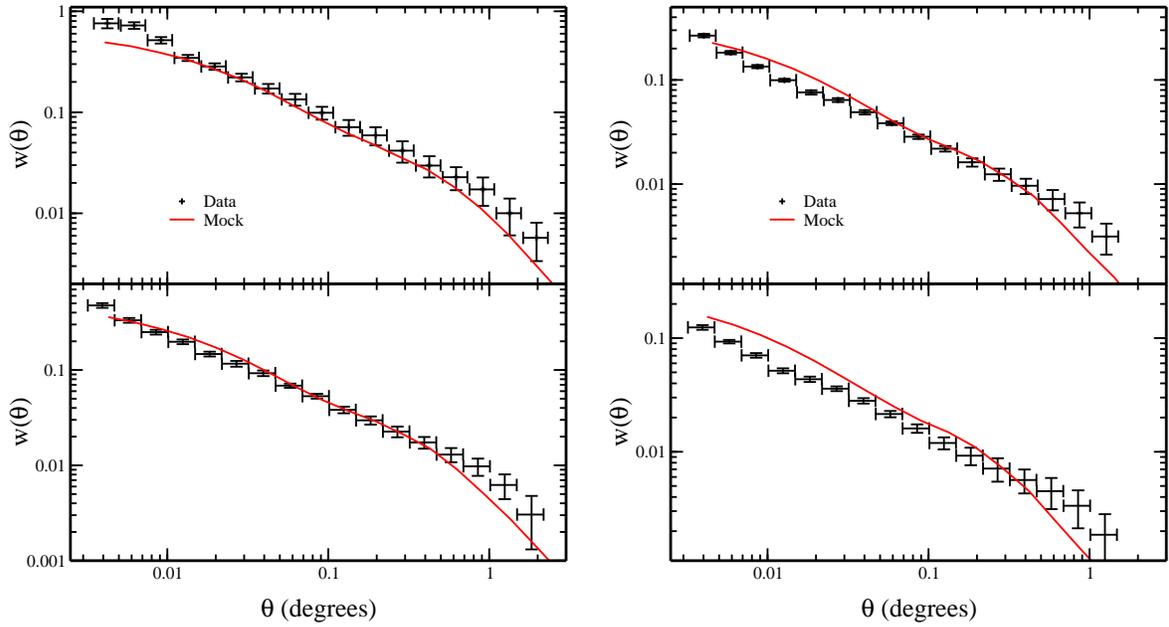}{f35.eps}
\caption{Comparisons of $w(\theta)$ measurements from mock catalogs and data.
In the left plot, the upper panel shows the comparison for 
$18 \le r^* \le 19$ and the lower panel for $19 \le r^* \le 20$.
The right plot does the same for $20 \le r^* \le 21$ (upper) and
$21 \le r^* \le 22$ (lower).  \label{fig:data_mock_comp}}
\end{figure}

In Figure~\ref{fig:data_mock_comp}, we see the direct comparison between the
data and the mock catalogs for the four magnitude bins.  The ``mock'' 
measurement is the mean value of $w(\theta)$ from 100 mock catalogs and the 
errors come from the diagonal elements of the covariance between these 100 
measurements.  In general, we see a reasonable agreement between the mock
catalogs and the data over the range of angular scales.  The one significant
discrepancy is in the small angle regime of the $21 \le r^* \le 22$ 
bin, where the mock catalog over-predicts the auto-correlation compared to 
our measurement.  This may be due to a slight error in the selection function 
for this magnitude bin or possibly some evolution in the biasing, but should 
not adversely affect the covariances from the mock catalogs.

	\subsection{Sub-Sampling Errors}

	In \S\ref{sec:cross}, we gave the formalism for calculating the 
cross-correlation errors from the variance of the mean measurement
determined in 35 non-overlapping square sub-samples of the data area.  This 
method works reasonably well for this purpose since we have a reasonable 
expectation that the sub-sample measurements are not strongly correlated.  
Since our primary estimators work on $\mathcal{O}(N^2)$ time, this method also
cuts down the processing time for the cross-correlation measurements by a 
factor of 35.

For the galaxy auto-correlation, we have mixing between physical scales due
to projection effects.  This means that we have correlations in the sub-samples
that are not modeled by Equation~\ref{eq:sub-error}, resulting in a likely  
under-estimation of the errors via this method.  

	\subsection{Jackknife Errors}\label{sec:jack-errors}

	In \S\ref{sec:large-scale}, we presented the formulae and method for 
calculating cross-correlation errors on large angular scales using the 
jackknife approach.  We can also apply this method to the calculation of the
galaxy auto-correlation errors on our full range of angular scales using the
estimator in Equation~\ref{eq:particle}.  In order to constrain the 26 
angular bins, we use 30 jackknife samplings of the data.  In principle, this 
should generate errors comparable to those from the simulations as this method 
should not suppress covariances on large scales.  In practice, however, the 
jackknife method appears to generate anti-correlations between angular scales 
smaller than the regions blocked out for each jackknife sample and the scale 
size of the blocked-out region (3 degrees in our implementation).

	\subsection{Gaussian Errors}
	
	At the large end of the angular spectrum, we can calculate the 
covariance matrix under the assumption that the error distribution is 
Gaussian in nature.  In this case, we will be using the pixelized version of 
the estimator for the auto-correlation (Equation~\ref{eq:pixel}).  The 
covariance for such an estimator is generically given by
\begin{eqnarray}
C(\theta_\alpha,\theta_\beta) &\equiv& \left \langle 
(\hat{w}(\theta_\alpha) - w(\theta_\alpha))
(\hat{w}(\theta_\beta) - w(\theta_\beta)) \right \rangle 
\nonumber \\
&=& \left \langle \hat{w}(\theta_\alpha) \hat{w}(\theta_\beta) \right \rangle 
- w(\theta_\alpha) w(\theta_\beta) \label{eq:gaussian-cov}
\end{eqnarray}
In order to calculate this, we need the covariance for $\delta$,
\begin{eqnarray}
C(\theta_{i,j}) &\equiv& 
\left \langle \delta_i \delta_j \right \rangle \nonumber \\
&=& w(\theta_{i,j}) + \frac{\delta_{ij}}{N_i},
\end{eqnarray}
where $\delta_{ik}$ is the traditional Kronecker delta, and 
$N_i$ is the number of galaxies in pixel $i$.  

Taking this into account for the estimator in Equation~\ref{eq:pixel}, the
covariance $C(\theta_\alpha,\theta_\beta)$ becomes
\begin{equation}
C(\theta_\alpha,\theta_\beta) = 2 \sum_{i,j} W_{ij,\alpha} \sum_{k,l} 
W_{kl,\beta} \left (w(\theta_{i,k}) + \frac{\delta_{ik}}{N_i} \right ) 
\left (w(\theta_{j,l}) + \frac{\delta_{jl}}{N_j} \right ),
\end{equation}
where we have dropped the fourth order terms which vanish under the assumption
that the over-densities have a Gaussian distribution. There are four terms to 
this sum, but we expect the cosmic variance term (the product of the two 
auto-correlations) to dominate on large scales, so we will calculate that to 
the exclusion of the others.  Inserting the form of the weight function, we 
must calculate
\begin{equation}
C(\theta_a,\theta_b) = 
\left (\frac{2}{\sum_{i^\prime,j^\prime} \Theta^a_{i^\prime,j^\prime}
\sum_{k^\prime,l^\prime} \Theta^b_{k^\prime,l^\prime}} \right )
\sum_{i,j,k,l} \Theta^a_{ij}\Theta^b_{kl} w(\theta_{i,k}) w(\theta_{j,l});
\end{equation}
a calculation which goes as the fourth power of the number of pixels (and is
the reason why we do this with pixels instead of by pairs).  

In this case both of our technical constraints push us in the same direction
toward larger angular scales.  First, we pay a heavy penalty for increasing 
the number of pixels in order to measure smaller and smaller angular scales.
Even a modest pixel size of one third of a degree will require on order 
$10^{13}$ computations.  At the same time, we know that the Gaussian 
approximation will break down at sufficiently small angles, so we have very 
little motivation for greatly increasing the number of pixels.  Indeed, the 
only reason to do so is to increase the range of angular scales where the 
Gaussian method reliably overlaps with the errors from the simulation method 
so as to allow for cross-checking.

	\section{Method Comparison}

	Despite the variations between the four error methods with respect to
applicable angular range and measurement technique, all of the methods produce
covariance matrices that have large off-diagonal elements as expected.  Moving
beyond this simple observation to a detailed comparison of the covariance
matrices for each of the four methods is not a trivial task.  As there is
no single well-established method for such a comparison, we present a number 
of tests to compare the shape and amplitude of the covariance matrices.  
Additional tests of the jackknife and sub-sample methods in comparison to 
simulated data sets can be found in Zehavi et al. (2001).  Due to the large 
difference in the angular range for the simulation, sub-sample
and jackknife errors and the Gaussian errors, we will set the latter aside 
initially and concentrate on comparing the first three methods.  We will 
finish by comparing the Gaussian covariance matrices to the appropriate parts 
of the covariance matrices from the simulations using the tools developed in 
the next sections.

	\subsection{Correlation Test}

\begin{figure}
\plotone{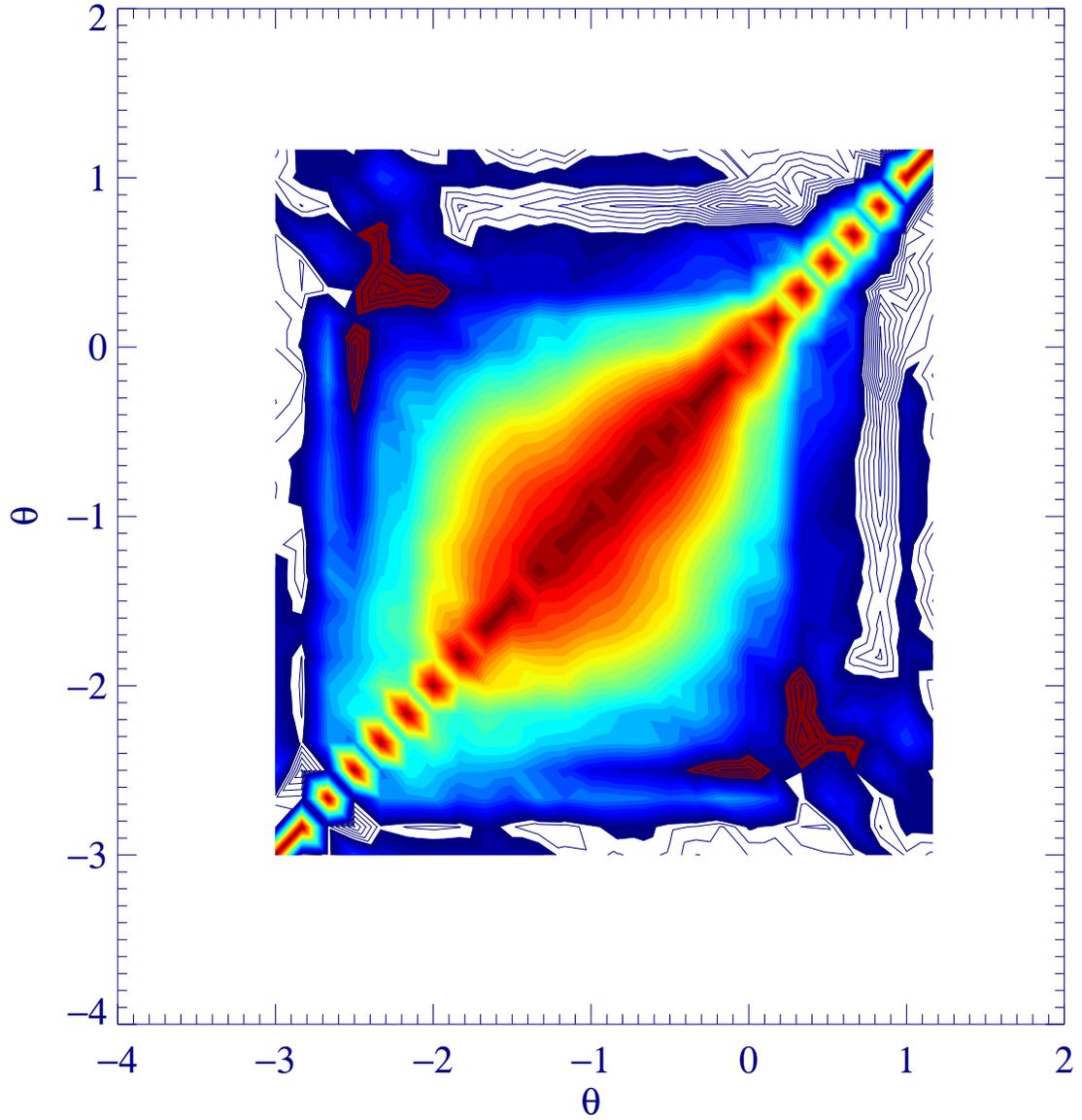}
\caption{Correlation matrix from simulations for the $21 \le r^* \le 22$
magnitude bin.  X and Y axes are the logarithms of the angular bins in 
degrees.  As predicted, the off-diagonal elements of the correlation matrix
are significant, regardless of the method used for calculating them.
\label{fig:mock_covar}}
\end{figure}

\begin{figure}
\plotone{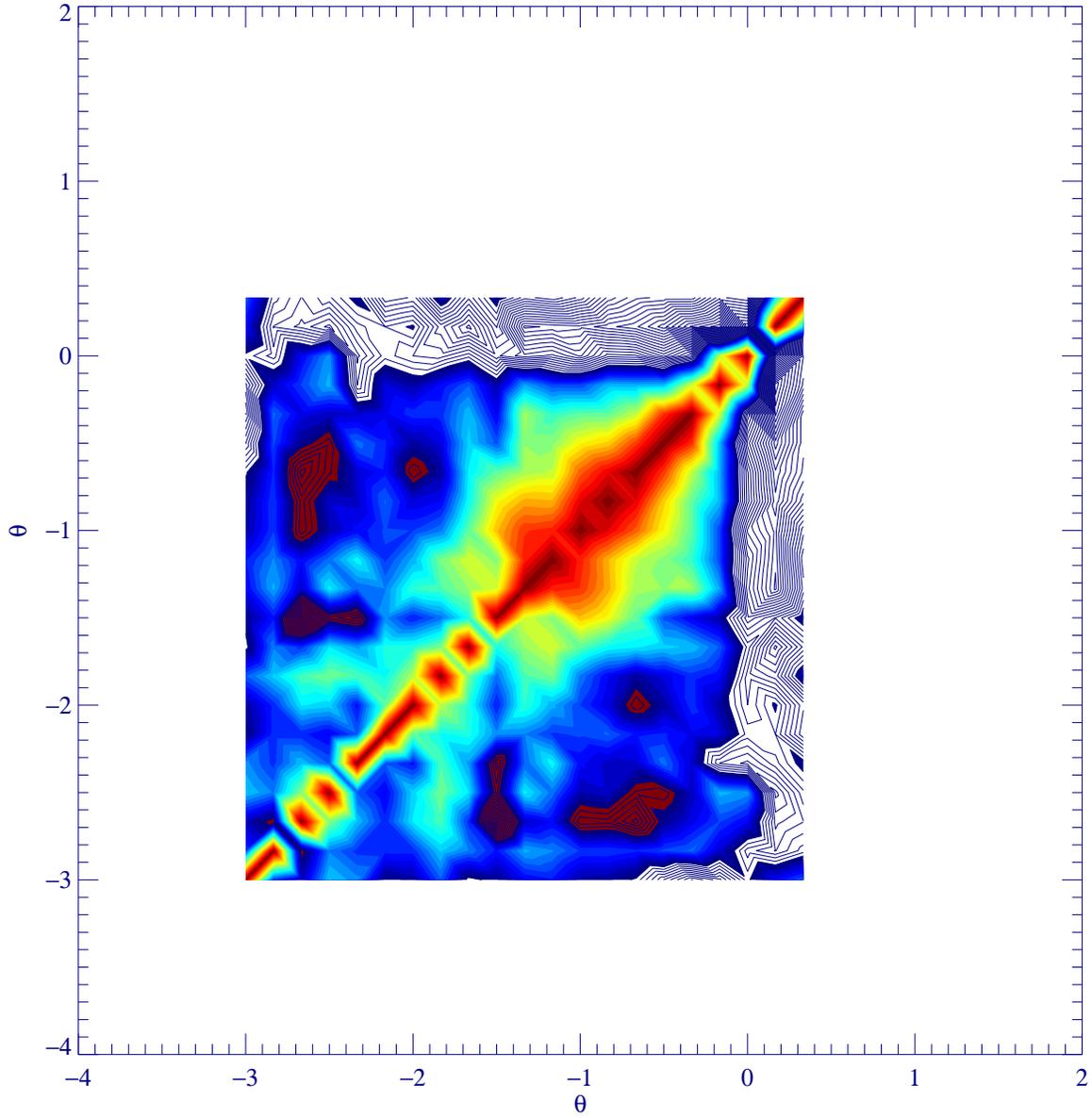}
\caption{Same as Figure~\ref{fig:mock_covar}, but for the sub-sample method.
\label{fig:sub_covar}}
\end{figure}

\begin{figure}
\plotone{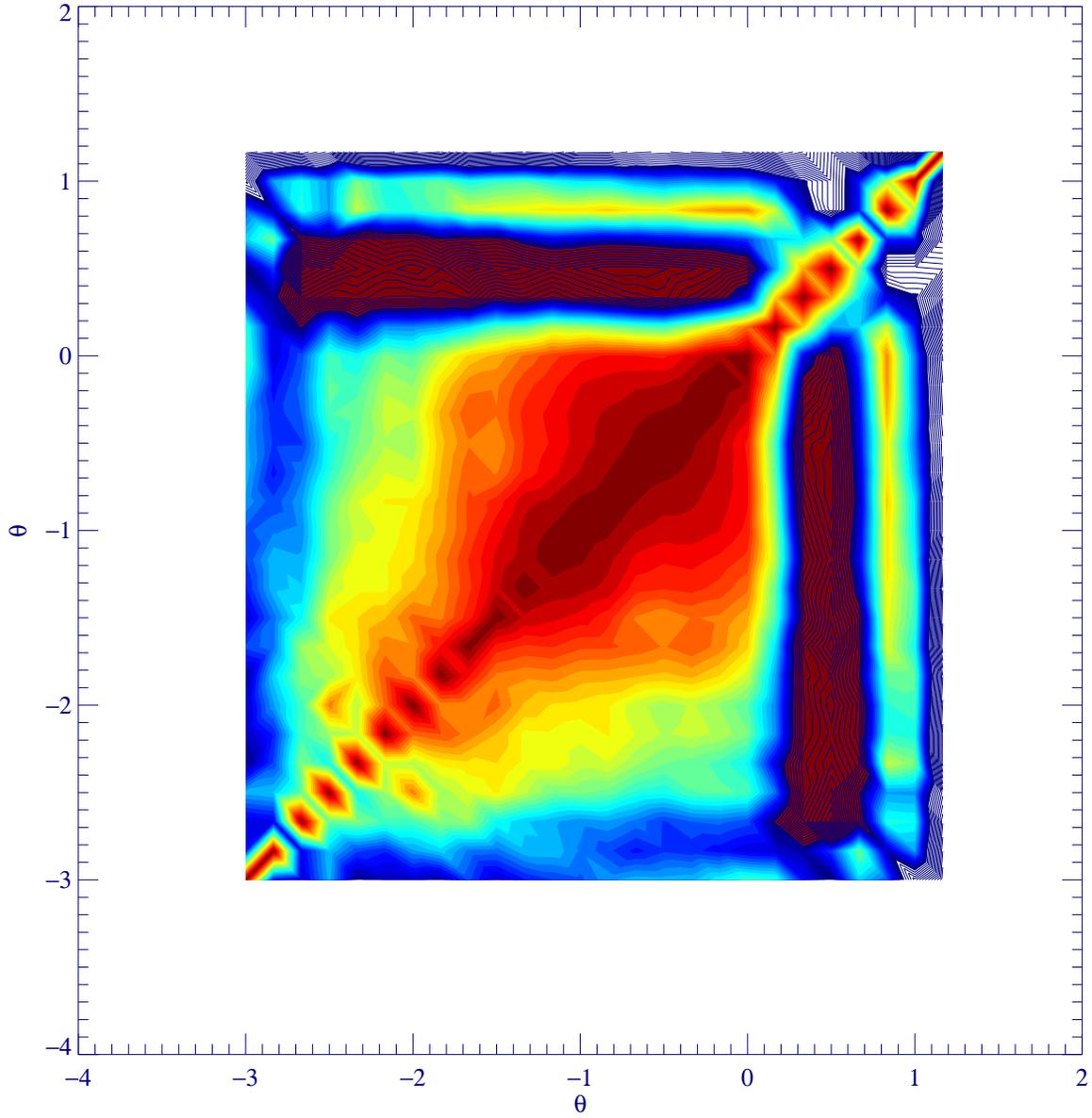}
\caption{Same as Figure~\ref{fig:mock_covar}, but for the jackknife method.
\label{fig:jack_covar}}
\end{figure}

Qualitatively, we can compare the shape of the covariance matrices by 
calculating the correlation matrices for each of the three main methods 
(simulations, jackknife, and sub-sample).  The elements of the correlation 
matrix ($r(\theta_\alpha,\theta_\beta)$) are given by 
\begin{equation}
r(\theta_\alpha,\theta_\beta) = \frac{C(\theta_\alpha,\theta_\beta)}
{\sqrt{C(\theta_\alpha,\theta_\alpha) C(\theta_\beta, \theta_\beta)}},
\end{equation}
where $r(\theta_\alpha,\theta_\alpha) \equiv 1$ and plotted for the three
methods in Figures~\ref{fig:mock_covar} through 
\ref{fig:jack_covar} for the $20 \le r^* \le 21$ magnitude bin.
No matter which set of matrices is to be used for further work (e.g. inverting 
$w(\theta)$ to recover the three dimensional power spectrum), it is clear 
that the full covariance matrix must be employed.  Likewise, we can see that
the sub-sample and jackknife errors have considerably stronger 
anti-correlations between the largest and smallest angular scales than 
the simulation correlations.  As discussed previously, this anti-correlation
for the sub-sample method is an artifact of the assumption that each 
sub-sample is independent of the others.  Likewise, for the jackknife errors,
the anti-correlation on the scale-size of the omitted regions strongly suggests
that the method is suppressing signal on that scale. 

	\subsection{Diagonal Test}

\begin{figure}
\plottwo{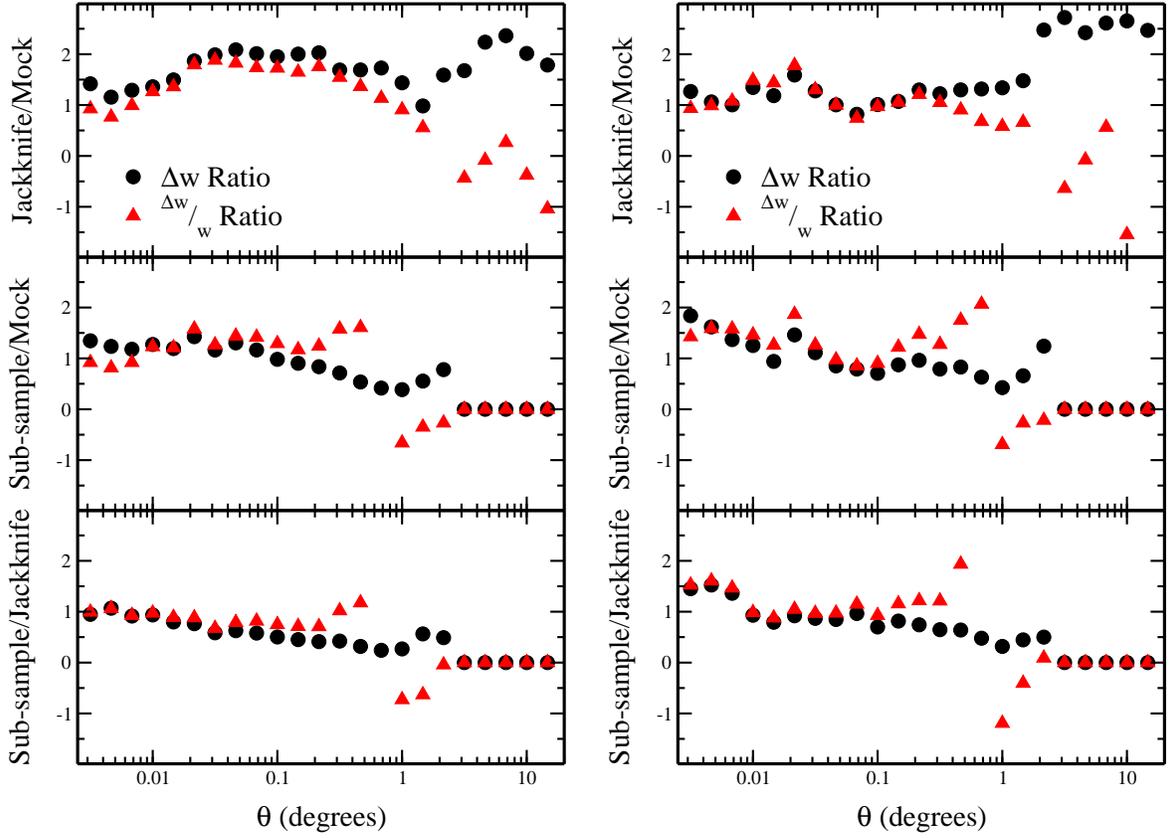}{f40.eps}
\caption{Ratios of $\Delta w$ and $\frac{\Delta w(\theta)}{w(\theta)}$ for 
the errors calculated using simulations, sub-sample and jackknife techniques 
for the $18 \le r^* \le 19$ (left) and $19 \le r^* \le 20$ (right) 
magnitude bins.\label{fig:error_ratio_bright}}
\end{figure}

\begin{figure}
\plottwo{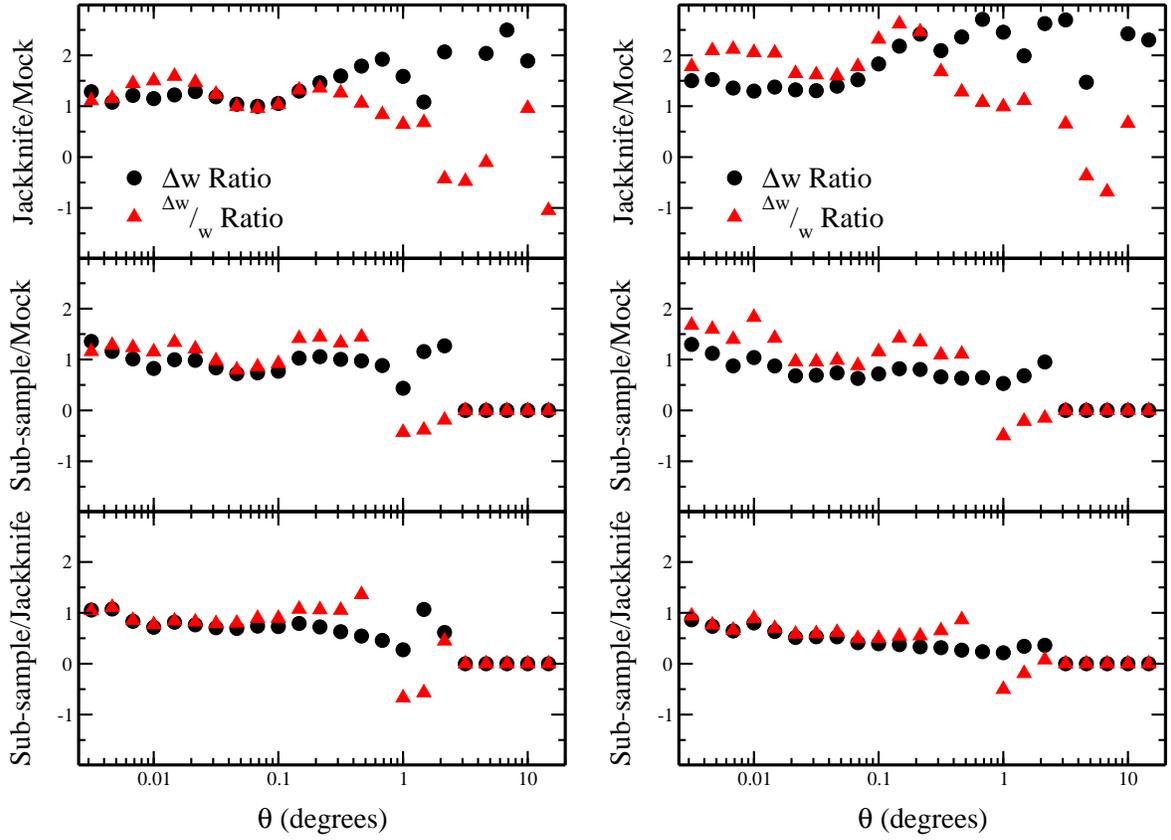}{f42.eps}
\caption{Same as Figure~\ref{fig:error_ratio_bright}, but for the
$20 \le r^* \le 21$ (left) and $21 \le r^* \le 22$ (right) 
magnitude bins.\label{fig:error_ratio_faint}}
\end{figure}

For a more quantitative comparison, we can examine the diagonal elements of
the covariance matrix.  Figures~\ref{fig:error_ratio_bright} and 
\ref{fig:error_ratio_faint} show the ratios of $\Delta w(\theta)$ and 
$\frac{\Delta w(\theta)}{w(\theta)}$ for the error calculations using the 
simulations, sub-samples and jackknife methods.  Note that, since we are 
including $w(\theta)$ from each of the methods in the second ratio, a 
negative value of $w(\theta)$ in only one of the methods can result in a 
negative ratio, while the first ratio is by definition positive.  The 
diagonals of the covariance matrices for the three methods generally agree to 
within factors of 1 to 2 for the three methods for the angular scales where 
they overlap, with considerable scatter as we approach the break in 
$w(\theta)$.  However, it is apparent that the $w(\theta)$ from sub-sample 
method starts to diverge from that measured from the simulation and jackknife 
methods at angular scales as small as $0.1^\circ$.  This makes the sub-sample 
method a questionable choice except at the smallest scales.

	\subsection{Product Test}

\begin{figure}
\plotone{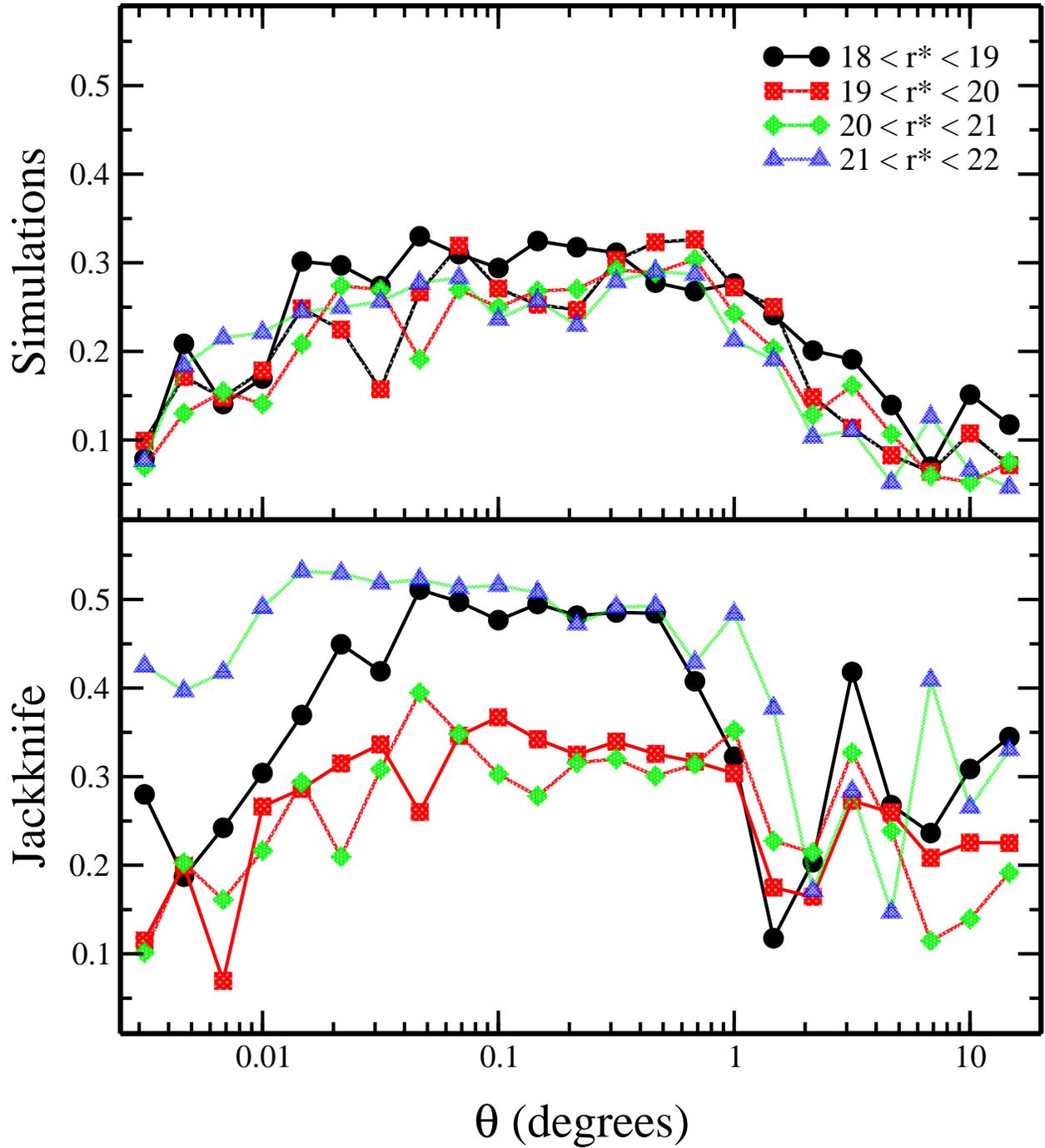}
\caption{$R(\theta)$ calculations for the simulations and jackknife correlation
matrices in all four magnitude bins. \label{fig:error_correlations}}
\end{figure}

For a more complete comparison of the simulation and jackknife covariance 
measures, we need to examine the off-diagonal elements.  To characterize the
contribution of the off-diagonal elements, we calculate $R(\theta)$, where
\begin{equation}
R(\theta_\alpha) \equiv \left ( \prod^N_\beta 
| r(\theta_\alpha, \theta_\beta)| \right )^{\frac{1}{N}}.
\end{equation}
For a perfectly diagonal matrix, this quantity will be zero and would be equal
to 1 for perfect correlation (or anti-correlation) between each angular bin.
Figure~\ref{fig:error_correlations} shows the values of $R(\theta)$ for each
of the methods in each of the magnitude bins.  As expected, both methods
show a fairly high degree of off-diagonal correlation by this measure, with
the jackknife generally showing larger amplitude off-diagonal elements than 
the mocks.  This confirms our earlier prediction that the jackknife errors 
are superior to other data-only measurements, but not as well-behaved as the
errors from simulations.

	\subsection{Wishart Likelihood Ratio Test}

Alternatively, we can attempt to further quantify the differences between the 
covariance matrices themselves by calculating their Wishart probabilities.  If
we measure the same quantity ($w(\theta)$, for instance) on several 
realizations of statistically similar data, then we expect the measurements
to be distributed according to the covariance matrix on that measurement.
Likewise, if we calculate the measurement covariance matrix for each 
realization of the data, then these covariance matrices will be distributed
around the ``true'' covariance matrix for the measurement and the data.  If 
we take the covariance matrix generated from the simulated catalogs to be the 
``true'' covariance matrix for data that is not contaminated by systematic 
errors, then the probability that the covariance matrix measured on one of
the simulated catalogs is drawn from a distribution of the true covariance 
matrix is given by the Wishart distribution (Wichern \& Johnson, 2002),
\begin{equation}
W(C_{S|J}|C_M) = \frac{|C_{S|J}|^{(N - N_\theta - 2)/2} 
e^{-tr(C_{S|J}C_M^{-1})/2}} {2^{N_\theta(N-1)/2} 
\pi^{N_\theta(N_\theta - 1)/4} |C_M|^{(N - 1)/2} \prod_{i=1}^{N_\theta} 
\Gamma((N - i)/2)}, \label{eq:wishart}
\end{equation}
where $C_{S|J}$ is the covariance matrix from either the sub-sample or 
jackknife methods, $C_M$ is the covariance matrix from the simulations, $N$ is 
the number of mock catalogs, $N_\theta$ is the number of angular bins.  If the
covariance matrix we test fit the distribution perfectly, this would
be equivalent to taking $W(C_M|C_M)$, which we denote as $W_0$.  With this in
hand, we can use Equation~\ref{eq:wishart} to perform a likelihood ratio test 
where
\begin{eqnarray}
q &=& \frac{W(C_{S|J})}{W_0} \nonumber \\  
\ln q &=& \ln W(C_{S|J}) - \ln W_0 \sim -\frac{1}{2}\chi_{S|J}^2,
\end{eqnarray}
where an acceptable $\chi_{S|J}^2$ would be roughly equal to the number of 
independent elements ($k$) of each covariance matrix: 
$k \equiv N_\theta (N_\theta + 1)/2$.

The application of this test assumes that $w(\theta)$ is distributed normally,
and thus the covariance matrix captures all of the information about the 
variation of $w(\theta)$.  While this is not likely exactly correct, we expect
it to be a reasonably good approxmation for the bulk of the distribution, with
perhaps some disagreement in the tails of the distribution.  The effect of this
is such that, for very small values of $W(C_{S|J}|C_M)$, we may not be able to 
accurately calculate the exact probability, but this should be sufficient for
our purposes.  In addition to these considerations, we also need to consider
the effect of disagreements between the amplitude of $w(\theta)$ in the 
simulated catalogs and the data (as shown in Figure~\ref{fig:data_mock_comp}).
To isolate the effects of the difference in methods and the difference in 
measurements, we must apply the sub-sample and jackknife methods to the 
simulated catalogs and define $C_S$ and $C_J$ as their respective means.  
Table~\ref{tab:wishart} gives the results of applying this test to covariance
matrices measured using the data and using the mean covariance matrices taken
from applying the methods to the set of simulations.  The first conclusion to 
draw from this calculation is that, although the deviation between the 
jackknife and sub-sample methods is not significant enough to choose between 
the methods using this test, both methods are reasonably acceptable 
alternatives to the simulation method if the latter is not available.  
Secondly, we can see from comparing the first and second sets of columns that 
the difference in the $w(\theta)$ for the real data and the simulation does
lead to some discrepancy between the apparent behavior of each of the 
methods and the behavior when one uses a common data set.

\begin{figure}
\plottwo{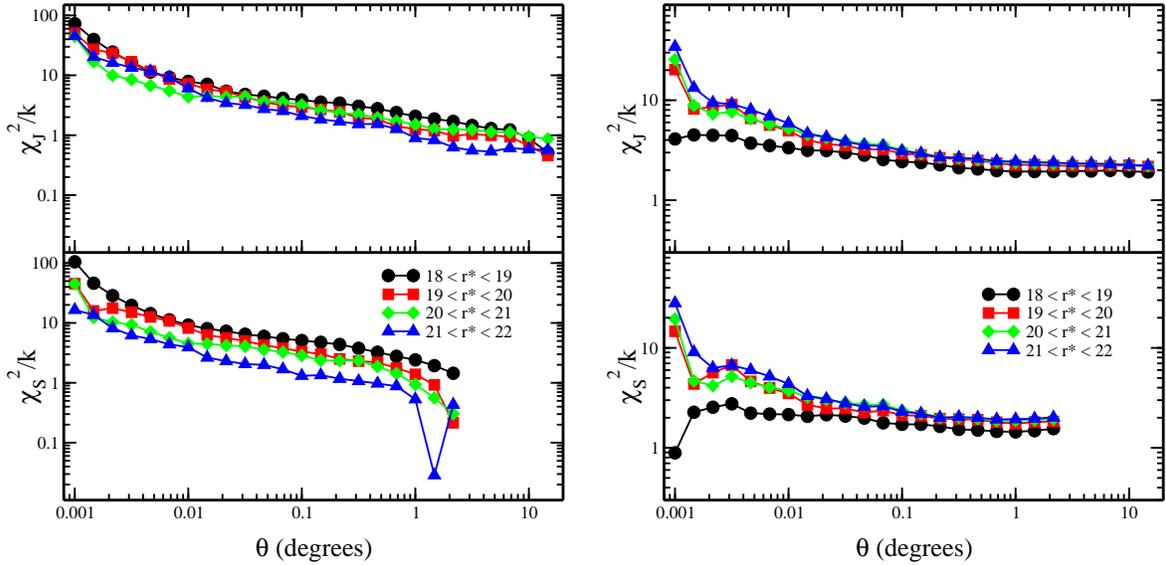}{f45.eps}
\caption{$\frac{\chi^2_{S|J}}{k}(\Theta)$ for data (left) and simulation 
(right) covariance measurements.\label{fig:wishart}}
\end{figure}

Finally, we can also generalize Equation~\ref{eq:wishart} to 
\begin{equation}
W(C_{S|J}(\Theta)|C_M(\Theta))= \frac{|C_{S|J}(\Theta)|^{(N - N_\Theta - 2)/2} 
e^{-tr(C_{S|J}(\Theta)(C_M(\Theta))^{-1})/2}} {2^{N_\Theta(N-1)/2} 
\pi^{N_\Theta(N_\Theta - 1)/4} |C_M(\Theta)|^{(N - 1)/2} 
\prod_{i=1}^{N_\Theta} \Gamma((N - i)/2)}, \label{eq:wishart_gen},
\end{equation}
where $C_{J|M}(\Theta)$ and $C_M(\Theta)$ are the sub-matrices corresponding
to the elements of the original covariance matrices where the angular bins 
involved are both are less than or equal to $\Theta$ and $N_\Theta$ is the 
corresponding number of angular bins.  Thus, calculating 
$\chi_{S|J}^2(\Theta)$ allows us to see how the results of the sub-sample and 
jackknife methods diverge from the simulation covariance as a function of 
angle.  As Figure~\ref{fig:wishart} shows, there is significant deviation 
at small angles from the relatively low values of $\chi_{S|J}/k$ given in 
Table~\ref{tab:wishart}, particularly for the covariance matrices from the data
itself.  This is likely due to the strong non-Gaussianity of the covariance 
matrices at small angles, an effect which would be diluted for larger 
$\Theta$.

	\subsection{Gaussian Comparison}

\begin{figure}
\plotone{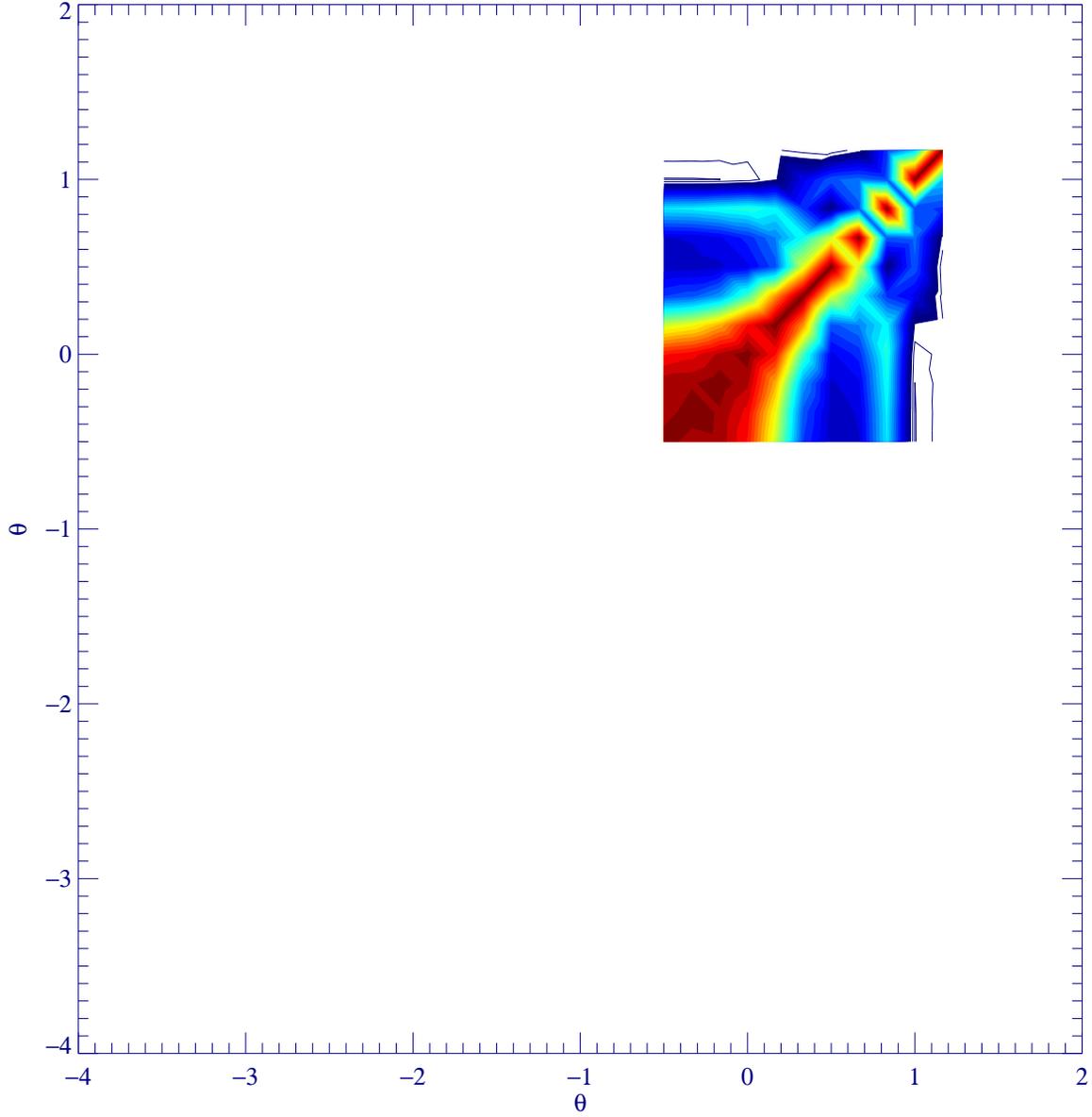}
\caption{Correlation matrix from Gaussian estimates for the 
$21 \le r^* \le 22$ magnitude bin.\label{fig:rect_covar}}
\end{figure}

\begin{figure}
\plottwo{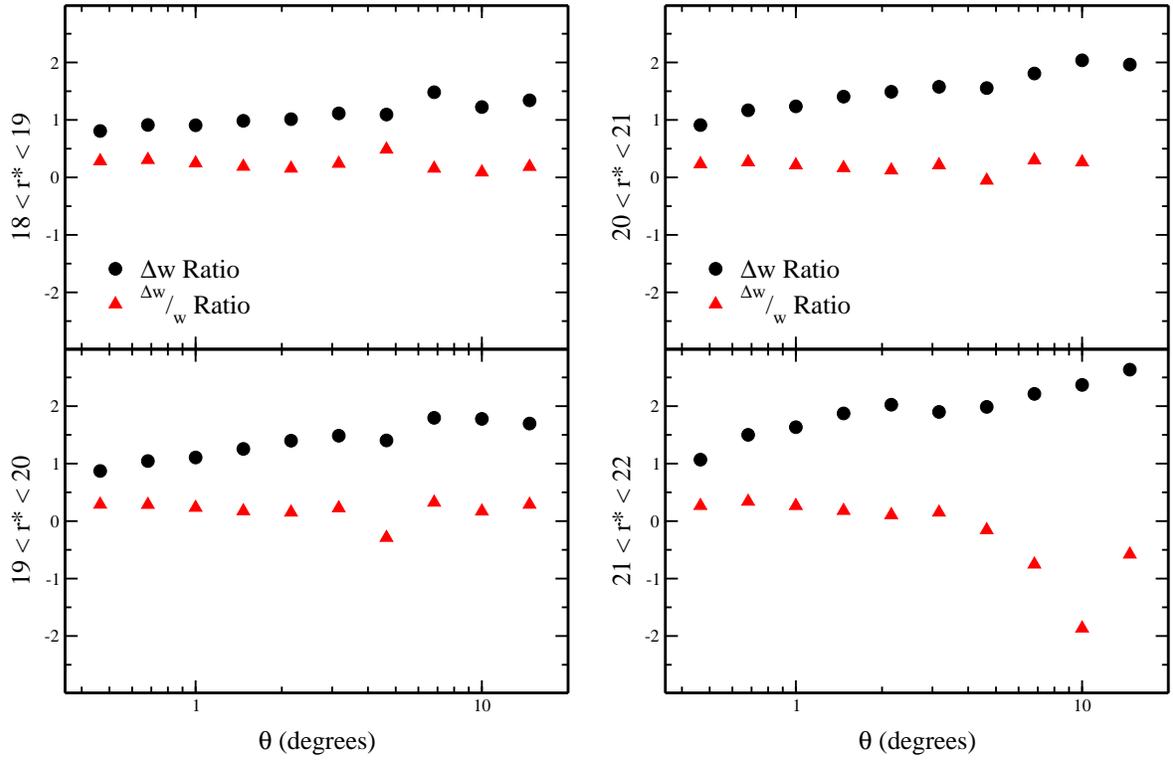}{f48.eps}
\caption{Ratios of $\Delta w$ and $\frac{\Delta w(\theta)}{w(\theta)}$ for 
the errors calculated using the Gaussian method compared to those found using
the simulation technique for the four magnitude bins.
\label{fig:error_large_ratio}}
\end{figure}

\begin{figure}
\plotone{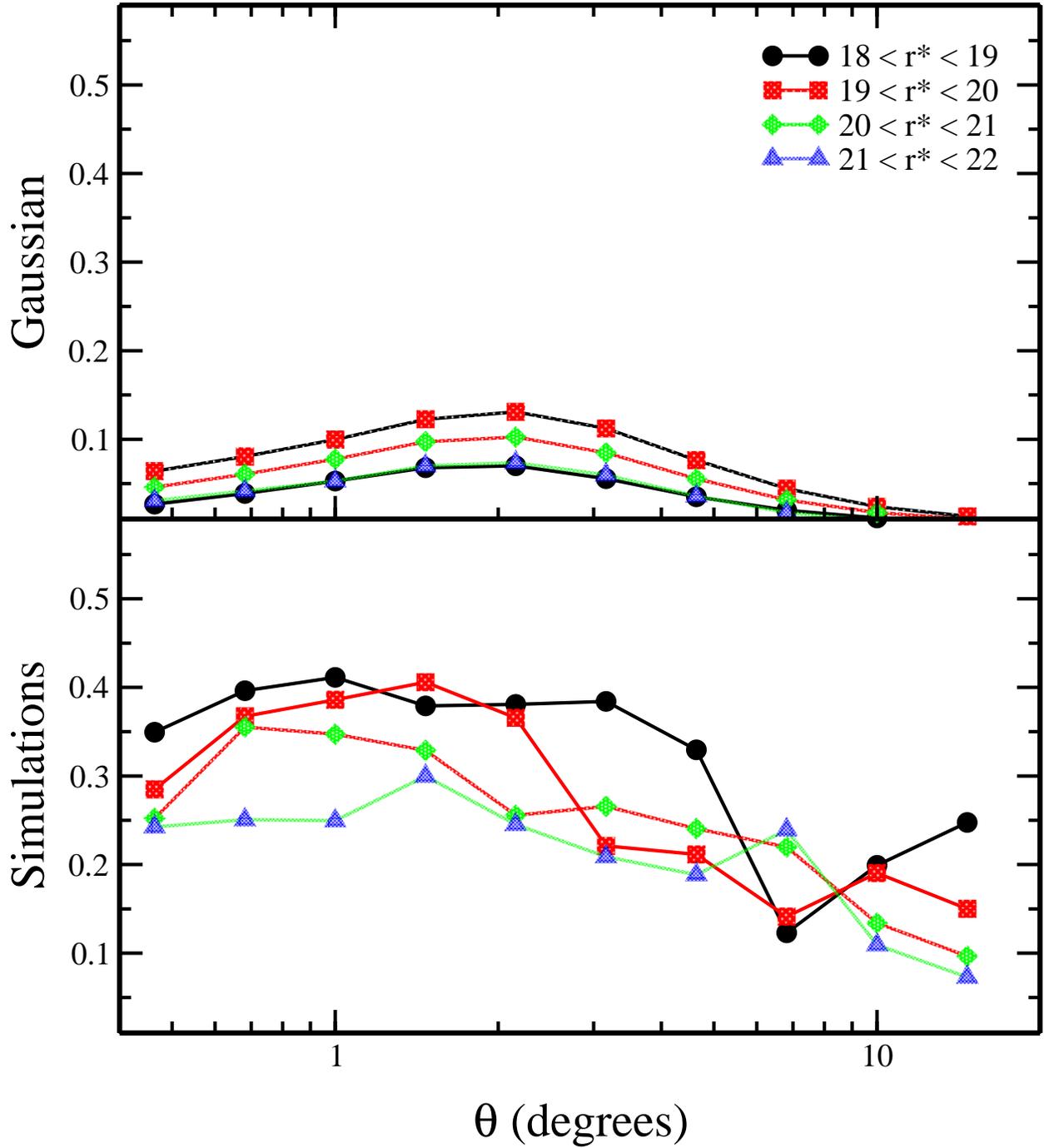}
\caption{$R(\theta)$ calculations for the Gaussian correlation matrices and the
corresponding elements of the simulation correlation matrices.
\label{fig:error_large_correlation}}
\end{figure}

We can use the same tools developed in the previous sections to analyze the 
agreement between the simulation errors and the Gaussian errors.
Figure~\ref{fig:rect_covar} shows the correlation matrix for the Gaussian 
method in the $21 \le r^* \le 22$ magnitude bin and 
Figure~\ref{fig:error_large_ratio} shows ratios of $\Delta w(\theta)$ and 
$\frac{\Delta w(\theta)}{w(\theta)}$ comparing the Gaussian errors to the
errors from the simulation.  Likewise, in 
Figure~\ref{fig:error_large_correlation}, we give the calculation for 
$R(\theta)$ for the Gaussian errors along with the $R(\theta)$ for the same
elements of the simulation correlation matrix.  As with the previous analysis,
the $\Delta w(\theta)$ ratios agree to within factors of 1-2, although the 
ratios of $\frac{\Delta w(\theta)}{w(\theta)}$ do demonstrate a clear 
difference in the large-angle behavior of $w(\theta)$ for the simulations and 
the data.  It is clear, however, that the disparity is due to different 
values for $w(\theta)$ for the two data sets and not due to a large 
disagreement between the errors.  For this reason (as well as the computational
difficulty of applying the Gaussian method to the set of simulated catalogs), 
we will forego the Wishart test for the Gaussian errors.

\section{Conclusions}

The SDSS will ultimately make definitive measurements of the angular 
clustering of galaxies.  To demonstrate the quality of the photometric data
produced by the SDSS, we studied two nights of commissioning data that
constitutes roughly a third of the Early Data Release.  We presented a means 
to extract galaxy probabilities from the output parameters of the photometric 
data processing pipeline.  With this in hand, we analyzed the effect of 
external factors (seeing variations, reddening extinction, stellar 
contamination and sky brightness) on that star/galaxy method.  After 
implementing cuts due to seeing, and to a lesser extent reddening and bright 
stars, we find no evidence for external systematic effects polluting the 
measurement of the angular correlation function.  

Having addressed external sources of error, we turned our attention to 
internal sources.  We investigated the possibility of differential 
photometric response across the CCDs as well as the introduction of 
correlations due to different photometric calibrations between the two runs
that made up our data area.  We verified that the $w(\theta)$ measurements
scaled according to Limber's equation in a $\Lambda$CDM cosmology and we 
were able to simulate possible deblending errors, placing a lower limit on the
angular extent of our clustering measurements.  As with the external 
systematic error checks, we were able to effectively eliminate systematic
errors after placing some constraints on the measurement.

Once the systematic errors were addressed for the broader category of 
angular measurements, we shifted our focus to the actual measurement of 
$w(\theta)$.  Using two complementary estimators, we were able to devise a
scheme for calculating $w(\theta)$ and its covariance using a reasonable 
amount of CPU time.  We presented a prescription for addressing the integral
constraint incurred due to a finite survey size and verified that the 
correction was much less than our measured $w(\theta)$.  Three methods were
discussed for calculating the covariance of $w(\theta)$: simulations, 
sub-samples and jackknife.  The covariances from the simulations proved most
reliable over the whole range of angular scales, but the jackknife errors 
would probably be adequate in the absence of simulations.  Finally, we 
compared the covariances from the simulations to estimates of the covariance
using a Gaussian assumption.  The Gaussian covariances generally had smaller
off-diagonal elements, but the diagonal elements agreed with the simulations
to within a factor of 2.

Companion papers will present the results for the angular correlation 
function $w(\theta)$ (Connolly et al. 2001) and the angular power spectrum, 
$C_l$ (Tegmark et al. 2001).  Using these measurements and estimates of the 
selection function, Dodelson et al. (2001) will extract the underlying 3D 
power spectrum and fit cosmological parameters.  A parallel analysis of the 
data using a KL decomposition (Szalay et al. 2001) will provide another 
set of parameter constraints.

Though they constitute only a small fraction of the data, the initial results 
strongly suggest that the remaining $98\%$ of SDSS photometric data will 
provide a powerful and robust set from which to gain cosmological information.

\section{Acknowledgements}

The Sloan Digital Sky Survey (SDSS) is a joint project of The University of 
Chicago, Fermilab, the Institute for Advanced Study, the Japan Participation 
Group, The Johns Hopkins University, the Max-Planck-Institute for Astronomy 
(MPIA), the Max-Planck-Institute for Astrophysics (MPA), New Mexico State 
University, Princeton University, the United States Naval Observatory, and the 
University of Washington. Apache Point Observatory, site of the SDSS
telescopes, is operated by the Astrophysical Research Consortium (ARC). 

Funding for the project has been provided by the Alfred P. Sloan Foundation, 
the SDSS member institutions, the National Aeronautics and Space 
Administration, the National Science Foundation, the U.S. Department of 
Energy, the Japanese Monbukagakusho, and the Max Planck Society. The SDSS Web 
site is http://www.sdss.org/. 

Support for this work was provided by the NSF through grant PHY-0079251 as 
well as by NASA through grant NAG 5-7092 and the DOE.

\clearpage

\include{bib}

\clearpage

\clearpage

\clearpage

\clearpage

\clearpage

\clearpage

\clearpage

\clearpage

\clearpage

\clearpage 

\begin{deluxetable}{ccccc}
\tablecaption{$\chi^2$ Values\label{tab:wishart}}
\tablewidth{0pt}
\tablehead{
\colhead{Magnitude} & \colhead{Data $\chi_J^2/k$} & 
\colhead{Data $\chi_S^2/k$} & \colhead{Simulation $\chi_J^2/k$}
& \colhead{Simulation $\chi_S^2/k$}
}
\startdata
$18 \le r^* \le 19$ & 0.51 & 1.4 & 1.9 & 1.6 \\
$19 \le r^* \le 20$ & 0.46 & 0.21 & 2.2 & 1.8 \\
$20 \le r^* \le 21$ & 0.87 & 0.3 & 2.2 & 1.9 \\
$21 \le r^* \le 22$ & 0.57 & 0.43 & 2.2 & 2.0 \\
\enddata
\tablecomments{The first two columns use the data-based jack-knife and 
sub-sample covariance matrices, respectively, to calculate $W(C_{S|J}|C_M)$ 
using Equation~\ref{eq:wishart}.  The third and fourth columns use covariance
matrices calculated by applying the jackknife and sub-sample methods to the 
simulated data to calculate $W(C_{S|J}|C_M)$.}
\end{deluxetable}

\end{document}

%% file: bib.tex
\section{References}
\def\refe {\par \hangindent=.7cm \hangafter=1 \noindent}
\def\apj { ApJ }
\def\astroph{{\tt astro-ph/}} 
\def\aap {A \& A }
\def\ajs{ ApJS }
\def\aj{AJ}
\def\prd{Phys ReV D}
\def\apjs{ ApJS }
\def\mnras { MNRAS }
\def\apjl { Ap. J. Let. }

\refe Bernstein, G.M. 1994, \apj, 424, 569
\refe Bouchet, F. R., Colombi, S., Hivon, E., \& Juszkiewicz, R.
	1995, \aap, 296, 575
\refe Buchert, T., Melott, A.L., \& Weiss, A. G. 1994, \aap, 288, 697
\refe Collins, C.A., Nichol, R.C., \& Lumsden, S.L. 1992, \mnras, 154, 295
\refe Connolly, A. et al., 2001, submitted to \apj, (\astroph0107417)
\refe Dodelson, S. et al., 2001, submitted to \apj, (\astroph0107421)
\refe Efstathiou, G. \& Moody, S. J. 2000, \astroph0010478
\refe J.H. Friedman, J.L. Bentley, \& R.A. Finkel 1977, ACM Transactions on 
Mathematical Software, 3(3):209
\refe Fukugita, M., Ichikawa, T., Gunn, J.E., Doi, M., Shimasaku, K., \& Scheider, D.P., 1996, \aj, 111,1748
\refe Fukugita, M., Yamashita, K., Takahara, F., Yoshii, Y., 1990, \apjl, 361, L1
\refe Gray, A. \& Moore, A. W. 2001, Proceedings of {\it Advances in 
Neural Information Processing Systems}, 13
\refe Gunn, J. E., \& The SDSS Collaboration, 1998,\aj, 116, 3040
\refe Hamilton, A.J.S. 1993, \apj, 417, 19
\refe Hui, L. \& Gaztanaga, E. 1999, \apj, 519, 622
\refe Huterer, D., Knox, L., \& Nichol, R. C. 2000, \astroph0011069
\refe Ivezic, et al., 2001, in preparation
\refe Kauffmann, G., Colberg, J.M., Diaferio, A. \& White, S.D.M. 1999, \mnras, 303, 188
\refe Landy, S. D, \& Szalay, A. S. 1993, \apj, 412, 64
\refe Lin, H., Yee, H. K. C., Carlberg, R. G., Morris, S. L., Sawicki, M., Patton, D. R., Wirth, G., \& Shepherd, C. W., 1999, \apj, 518, 533
\refe Lupton, et al., 2001, in preparation
\refe Maddox, S.~J., Efstathiou, G., Sutherland, W.~J., \& Loveday, L. 1990, \mnras, 242, 43P
\refe Maddox, S.J. Maddox, Efstathiou, G., Sutherland, W.J., 1996, \mnras, 283, 1227
\refe Moutarde, F., Alimi, J.-M., Bouchet, F.R., Pellat, R., \& Ramani, A. 1991,\apj, 382, 377
\refe Navarro, J.F., Frenk, C.S. \& White, S.D.M. 1997, \apj, 490, 493
\refe Peacock, J. A. \& Dodds, S. J. 1994, \mnras, 267, 1020
\refe Peebles, P.J.E. 1980, {\it The Large Scale Structure of the Universe} 
	(Princeton University Press)
\refe Petrosian, V., 1976, \apj, 209, L1
\refe Schlegel, D.J., Finkbeinder, D. P., \& Davis, M. 1998, \apj, 500, 525
\refe Scoccimarro, R. 2000, \apj, 544, 597
\refe Scoccimarro, R., Sheth, R.K., Hui, L. \& Jain, B. 2001, \apj, 546, 20
\refe Scoccimarro, R. \& Sheth, R.K. 2001, astro-ph/0106120
\refe Sheth, R.K. \& Lemson, G. 1999, \mnras, 305, 946
\refe Sheth, R.K. \& Diaferio, A. 2001, \mnras, 322, 901
\refe Szalay, A.S. et al., 2001, submitted to \apj, (\astroph0107419) 
\refe Szapudi, S., Colombi, S. \& Bernardeau, F. 1999, \mnras, 310, 428
\refe Szapudi, S., Szalay, A. S. 1998, \apjl, 494, L41 
\refe Tegmark, M. et al., 2001, submitted to \apj, (\astroph0107418) 
\refe Tegmark, M., Hamilton, A. J. S., Strauss, M. A., Vogeley, M. S., 
\& Szalay, A. S. 1998, \apj, 499, 555
\refe Wichern, D.W. \& Johnson, R.A., 2002, {\it Applied Multivariate 
Statistical Analysis}, ${\rm 5^{th}}$ Edition, (Prentice Hall)
\refe Yasuda, N.~et al.\ 2001, \aj, 122, 1104 
\refe York, D. G., \& The SDSS Collaboration 2000, \aj 120, 157
\refe Zehavi, I. et al.\ 2002, \apj, 571, 172